\documentclass[%
reprint,
superscriptaddress,
prd,
floatfix,
]{revtex4-2}

\usepackage{physics}

\usepackage{graphicx}
\usepackage[caption=false]{subfig}
\usepackage{hyperref}
\usepackage{cleveref}
\usepackage{dcolumn}
\usepackage{booktabs}
\usepackage{float}
\usepackage{lipsum}

\usepackage[english]{babel}
\usepackage[T1]{fontenc}
\usepackage{dsfont}
\usepackage[mathlines]{lineno}
\usepackage{braket}
\usepackage{nicefrac}

\usepackage[export]{adjustbox}
\usepackage{xcolor, etoolbox}

\usepackage{tikz}
\usetikzlibrary{arrows.meta, positioning}

\usepackage{fontspec}
\usepackage{unicode-math}
\definecolor{bordeauxRed}{RGB}{178, 34, 52}
\definecolor{royalBlue}{RGB}{65,105,225}
\definecolor{oliveGreen}{RGB}{128,128,0}

\hypersetup{
    colorlinks,
    linkcolor=bordeauxRed,
    citecolor=royalBlue,
    urlcolor=oliveGreen,
    breaklinks=true 
}

\crefname{figure}{Fig.}{Figs.}
\Crefname{figure}{Fig.}{Figs.}
\crefname{section}{Sec.}{Secs.}
\Crefname{section}{Sec.}{Secs.}
\crefname{appendix}{Appendix}{Appendices}
\Crefname{appendix}{Appendix}{Appendices}
\crefname{equation}{Eq.}{Eqs.}
\Crefname{equation}{Eq.}{Eqs.}
\creflabelformat{equation}{#2\textup{(#1)}#3}

\begin{document}
\title{Inverse Scattering \& Isospectrality in (Analog) Cosmology}

\author{Christian F. Schmidt}
\affiliation{Theoretisch-Physikalisches Institut, Friedrich-Schiller-Universität Jena, Max-Wien-Platz 1, 07743 Jena, Germany}

\author{Stefan Floerchinger}
\affiliation{Theoretisch-Physikalisches Institut, Friedrich-Schiller-Universität Jena, Max-Wien-Platz 1, 07743 Jena, Germany}

\begin{abstract}
Cosmological particle production — the excitation of field quanta by a dynamical spacetime — underlies the primordial perturbations observed in the cosmic microwave background and has been realized in analog quantum simulators with ultra-cold atoms. 
We ask the inverse question: given the excitation spectrum, can the cosmological history that produced it be reconstructed, and is that reconstruction unique?

We approach this problem through a mapping of particle production to one-dimensional quantum scattering, in which the expansion history plays the role of a scattering landscape and the produced spectrum the role of reflected and transmitted waves. 
This mapping reveals a rich interplay between scattering theory and cosmology. 
The scale factor enters as a zero-energy resonance — an exceptional case of scattering theory — while cosmic singularities correspond to bound states, so that regular cosmologies are governed entirely by the scattering continuum. 
We find an interesting isospectrality: histories related by an inversion of the scale factor (expansion ↔ contraction) yield identical particle numbers, differing only by a $\pi$-shift in phase.
Given both the particle number (or alternatively the amplitude) and phase of the cosmological power spectrum, the reconstruction is unique for regular, asymptotically static cosmologies.

We demonstrate the reconstruction both analytically, for exactly solvable scattering landscapes, and numerically, recovering common logistic expansion histories to better than $0.02 \%$.
We further quantify its robustness to the momentum-space resolution and measurement uncertainties characteristic of analog quantum simulation. 
This work establishes a direct link between the observable power spectrum and the cosmological history that produced it, offering a new perspective on the classification of early-universe scenarios and the construction of alternatives to inflation.
\end{abstract}

\maketitle
\section{Introduction}
The inversion of wave scattering has a longstanding history in physics (which is extensively summarized in the foreword of \cite{Chadan1989}), beginning with Rayleigh's discussion on the possibility to recover a variable density profile from the frequencies of acoustic vibrations that are sourced by it \cite{Rayleigh1877} —
a question that has been famously phrased as "Can one hear the shape of a drum?" by Kac \cite{Kac1966}.
With the advent of quantum mechanics as a theory of matterwaves, the inverse problems attracted mathematicians who sought for a spectral reconstruction of the general Sturm-Liouville problem \cite{Ambarzumian1929, Borg1946, Krein1951, LevitanGasymov1964} while 
physicists consulted exactly solvable models to gain information on interatomic potentials via the molecular spectra \cite{RosenMorse1932,Poeschl1933}.

Studies on the uniqueness of solutions to the inversion of the three-dimensional Schrödinger equation in a central potential \cite{Bargmann1949a,Bargmann1949b,Levinson1949} were followed by a uniqueness theorem of Marchenko \cite{Marchenko1950},
and shortly afterwards by the celebrated constructive method of Gelfand \& Levitan \cite{GelfandLevitan1951},
which was further developed by Jost \& Kohn \cite{JostKohn1953} as well as Levinson \cite{Levinson1953}.
An alternative route was taken in the meantime by Marchenko \cite{Marchenko1952, Marchenko1953, Marchenko1955} based on concepts introduced by Jost \cite{Jost1947}.
Both methods culminated in the so-called Gelfand-Levitan-Marchenko equations \cite{Dunajski2010}.
A systematic theory of the one-dimensional inverse scattering problem, which is of particular relevance for this work, has been developed by Faddeev \cite{Faddeev1958,Faddeev1964,Faddeev1976} based on Marchenko's concepts
and subsequently refined by Deift \& Trubowitz \cite{DeiftTrubowitz1979} and Newton \cite{Newton1980}. 
We refer to the resulting set of equations as the Faddeev-Marchenko equations (FMEs), to avoid the common confusion with the Gelfand-Levitan route (see \cite{Burridge1980,Newton1980} for a clarification).
Beyond a spectral reconstruction, these concepts also serve as a design principle: Kay \& Moses used them to build the class of potentials that transmit
without reflection at any energy \cite{KayMoses1956a}, an early instance of shaping a scattering landscape to produce a desired propagation of waves.

The reach of these equations extends well beyond quantum mechanics, since the stationary Schrödinger equation is also the equation of a wave at normal incidence on a stratified medium. 
Recovering a medium from the signal it reflects is a common interest in the practical fields of seismic exploration, ultrasonic and acoustic probing, and of the characterization of layered dielectrics \cite{WareAki1968, Claerbout1968, GopinathSondhi1970, Kay1972, DeiftTrubowitz1979, Burridge1980, BerrymanGreene1980, Coen1981}; 
the reflectionless landscapes just mentioned were in fact constructed for dielectric profiles \cite{KayMoses1956a}.
In these practical fields, it is more common to reconstruct the wavefield rather than the medium \cite{Rose2002, BrogginiSnieder2012, Wapenaar2014, Wapenaar2021,Oyekan2026} which is also the reconstruction principle employed in this work.

The same inversion methods reappear as a pivotal step of the inverse scattering transform,
which solves integrable nonlinear partial differential equations such as the Korteweg--de Vries
equation, with the reflectionless case yielding their soliton solutions \cite{Gardner1967, Lax1968, Gardner1974, Caudrey1976, Dunajski2010}.
The construction extends to a large class of nonlinear, integrable systems \cite{Dunajski2010,NeugebauerMeinel2003}, including those of general relativity, where Neugebauer \& Meinel constructed the rigidly rotating disk of dust as an exact solution to Einstein's field equations \cite{NeugebauerMeinel1993,NeugebauerMeinel1994,NeugebauerMeinel1995}.

Acoustic phenomena nowadays feature prominently as simulators for curved spacetime kinematics as a substantial part of the analog gravity program (see \cite{Schuetzhold2025,BarceloLiberatiVisser2026} for recent reviews).
Therein, acoustic spacetimes emerge on large scales and a space- and time-dependent speed of sound can be engineered to let the sound-waves effectively experience a dynamical cosmos, a black- or white hole, or be subjected to superradiant scattering via a draining-bathtub geometry.  
This renders phenomena of quantum field theory in curved spacetime \cite{BirrellDavies1982,mukhanov_winitzki_2007} experimentally observable.
For instance, cosmological particle production \cite{Parker1969} — the mechanism held responsible for the primordial perturbations whose imprint is observed in the cosmic microwave background \cite{MukhanovChibisov1981, MukhanovFeldmanBrandenberger1992, Planck2018PowerSpectrum} —
has by now been observed on several of these platforms \cite{Hung2013, Eckel2018, Chen2021, Steinhauer2022, Viermann2022, Sparn2024}.

Among them, the quantum field simulator realized in a Bose-Einstein condensate with time-dependent interactions \cite{Viermann2022, Tolosa2022}, 
the rate of expansion is set by the interaction strength between the atoms, which is adjusted through an external magnetic field, so that an entire family of cosmological histories can be programmed and repeated at will.
The quantum simulation returns a phononic excitation power spectrum, extracted \emph{in-situ} from correlations of the density fluctuations \cite{Viermann2022, Sparn2024}.
In earlier work \cite{Schmidt2024,Sparn2024} we have shown that the evolution of the field modes in such a simulator is equivalent to a stationary scattering problem in one spatial dimension (the sound-horizon) in which the expansion history plays the role of a scattering landscape and the produced spectrum the role of the reflected and transmitted waves (cf.\@ \cite{Sparn2024} for experimental results).
It has since been extended to dispersive backgrounds \cite{Schmidt2026} and to further simulation platforms \cite{Schmidt2025PRA}.

In the present work we ask the converse question: given the measured spectrum, can the history that produced it be reconstructed, and is that reconstruction unique? 
Or adopting the phrasing of Kac \cite{Kac1966}: "can one hear the shape of an acoustic spacetime?" 
We believe this question to be natural, given the firm root of the analog simulators in acoustics.
Due to the structural analogy, we automatically ask the same question in the cosmological target system, namely to what extent the history of the early universe is determined by the spectrum it left behind.

\emph{The remainder of this paper is organized as follows.}
In Sec.~\ref{sec:CosmoScattering}, we recall the formulation of cosmological particle production and one-dimensional quantum scattering,
discuss the role of spacetime singularities and showcase an expansion-contraction duality via a Darboux-transformation.
In Sec.~\ref{sec:InverseScattering} we state the inverse problem, solve it in closed form for two unimodal scattering landscapes and use them as a benchmark for our numerical solver.
In Sec.~\ref{sec:NumericalReconstruction}, we numerically solve the inverse problem for two expansion histories that are either well-known in quantum field theory in curved spacetime (Bernard-Duncan model) or analog quantum simulation (quench) 
by recovering both the scale factor and the underlying scattering landscape from their spectra alone.
The same section assesses the robustness of the procedure to momentum space resolution limits and measurement errors in the power spectrum.
Finally, we summarize our findings and provide an outlook in Sec.~\ref{sec:Conclusion}.

\section{Cosmological Particle Production as One-Dimensional Scattering}
\label{sec:CosmoScattering}
We briefly summarize the key concepts and elaborate further on the role of singularities and isospectral scenarios.

\subsection{The Scattering Analogy}
\begin{figure}
\centering
    \includegraphics[width=\columnwidth]{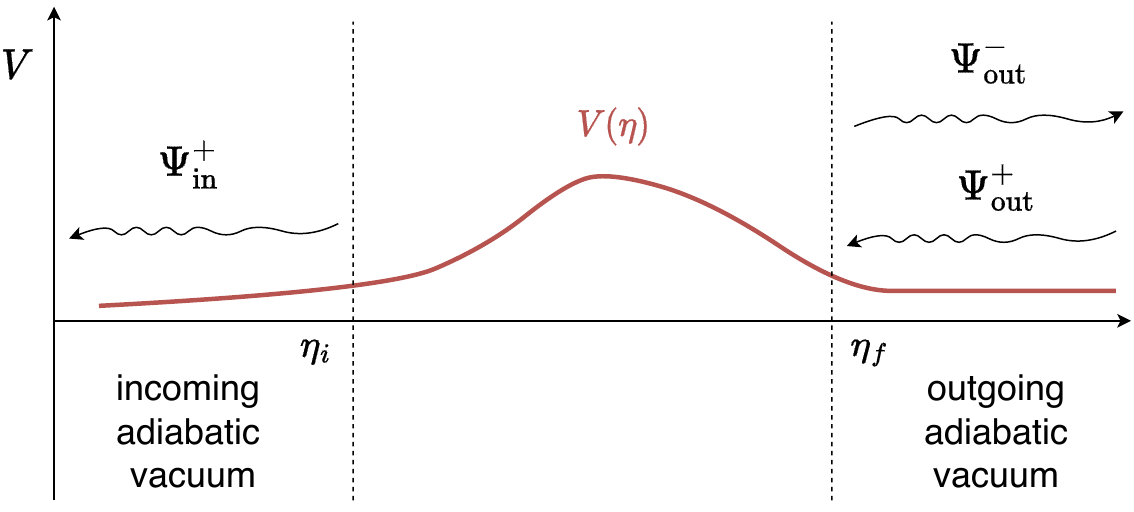}
    \caption{Scattering analogy in asymptotically static spacetimes: An initial vacuum defined by the positive frequency mode $\Psi^+_\mathrm{in}$ is transferred into a final vacuum state associated to the positive frequency mode $\Psi^+_\mathrm{out}$, which is 
    non-empty due to the presence of the negative-frequency mode $\Psi^-_\mathrm{out}$.}
    \label{fig:ScatteringAnalogyWKB}
\end{figure}
We investigate a massless scalar field $\phi$ which is minimally coupled to a cosmological spacetime with metric $g_{\mu \nu}(x)$
in terms of the effective action 
\begin{equation}
    \Gamma[\phi] = - \frac{1}{2} \int \mathrm{d}^{D+1} x \sqrt{g} g^{\mu \nu} \partial_\mu \phi \partial_\nu \phi
\end{equation}
The metric $g_{\mu \nu}(x)$ is associated to the Friedmann-Lemaître-Robertson-Walker (FLRW) line-element is $\mathrm{d}s^2 =  [- \mathrm{d}t^2 + a^2(t) \mathrm{d}\Omega_D^2]$
with the scalefactor $a(t)$ and the induced spatial line-element $\mathrm{d}\Omega_D^2$. The Jacobian determinant is denoted as $\sqrt{g} \equiv [- \det g_{\mu \nu}]^{1/2}$.
It is common to transform to the conformal time coordinate $\mathrm{d}\eta = \mathrm{d}t / a(t)$ and to rescale the field $\phi$ by the conformal factor $a^w(\eta)$, i.e.\@ to introduce 
the re-scaled field 
\begin{equation}
    \chi(\eta,\mathbf{x}) = a^w(\eta) \, \phi(t(\eta),\mathbf{x}), \quad w = (D-1) / 2,
    \label{eq:FieldRedefinition}
\end{equation}
where $w$ is called the conformal weight of the scalar field. 
The conformal factor follows from the transformation behavior of $\phi$ under the conformal group \cite{BirrellDavies1982}.
Quantizing $\chi$ leads to a mode equation of the well-known form
\begin{equation}
    \left[ -\dv[2]{\eta}  + V(\eta) \right] \psi(k, \eta) = - h(k) \psi(k, \eta)
    \label{eq:ModeEquation}
\end{equation}
with the scattering potential 
\begin{equation}
    \begin{aligned}
        V(\eta) = \frac{D-1}{2} \left[\frac{a''(\eta)}{a(\eta)} - \frac{3-D}{2}\left( \frac{a'(\eta)}{a(\eta)} \right)^2\right],
        \label{eq:ScatteringPot}
    \end{aligned}
\end{equation}
and the eigenvalue
\begin{equation}
    h(k) = \begin{cases} -k [k + (D-1) \sqrt{\abs{\kappa_\mathrm{sp}}}] &\text{for} \quad \kappa_\mathrm{sp} > 0, \\
        -k^2 &\text{for} \quad \kappa_\mathrm{sp} = 0, \\
        - [k^2 + \abs{\kappa_\mathrm{sp}} (D-1)^2/4 ] &\text{for} \quad \kappa_\mathrm{sp} < 0, \end{cases}
        \label{eq:LaplacianEigenvalue}
\end{equation}
which depends on the normalized spatial curvature constant $\kappa_\mathrm{sp}$.

If the cosmological spacetime is asymptotically static, i.e. the scale factor becomes constant as $\eta \to \pm \infty$, the scattering potential decays to zero towards future- and past-infinity where we can impose an adiabatic vacuum (cf.\@ \cref{fig:ScatteringAnalogyWKB}).
The key difference between the cosmological and quantum simulation setup is that in the latter the scale-factor $a$ can be kept constant initially and finally.
Then the potential in \cref{fig:ScatteringAnalogyWKB} has a compact support, and unique initial and final vacua exist (cf.\@ \cite{Schmidt2024} for details).
In the cosmological case, by contrast, the adiabatic vacuum is one choice among many (although a well-established one \cite{Fulling1989}).
The term cosmological particle production then refers to the pair-creation process due to these dynamics which lead to a two-mode squeezed state.
The non-emptiness of the outgoing adiabatic vacuum (or ground-state of the final Hamiltonian) with respect to the initial particle notion is described by reflection in the scattering analogy.
As we discussed in detail in \cite{Schmidt2024}, the final vacuum spectrum can be expressed in terms of scattering amplitudes $r_+(k)$ and $t(k)$.
In terms of a (quasi)-particle annihilation operator $a_\mathbf{k}$, the particle number is
\begin{equation}
    N_k \equiv \langle a^\dagger_\mathbf{k} a_\mathbf{k} \rangle = \bigg \lvert \frac{r_+(k)}{t(k)} \bigg \rvert^2,
    \label{eq:ParticleNumber}
\end{equation}
where $r_+(k)$ is the amplitude of a wave reflected into the direction of positive $\eta$ and whose absolute value sets the squeezing strength.
The particle pairs interfere according to their coherence $\Delta N_k = \mathrm{Re}[\langle a_\mathbf{k} a_{-\mathbf{k}} \rangle]$, which in the final static epoch oscillates as
\begin{equation}
    \Delta N_k = \bigg\lvert \frac{r_+(k)}{t^2(k)} \bigg\rvert \cos\big(2\sqrt{-h(k)}\,\eta + \vartheta_k\big), \quad \vartheta_k = \arg[-r_+(k)],
    \label{eq:Amplitude}
\end{equation}
where the phase $\vartheta_k$ is defined relative to the origin of conformal time and is fixed by the measured signal.
The main point here is that such output states can be experimentally investigated in modern quantum simulations with ultracold atoms. 
In particular these provide direct access to $r_+(k)$.

\subsection{The Role of Cosmic Singularities}

\subsubsection{Zero-Energy Resonance}
A peculiar property of the scattering potentials is that their definition \eqref{eq:ScatteringPot} is equivalent to the existence of a zero-energy solution to \eqref{eq:ModeEquation},
\begin{equation}
    \left[-\dv[2]{\eta} + V(\eta)\right] a^w(\eta) = 0,
    \label{eq:ZeroEnergyResonance}
\end{equation}
which is identical to the conformal factor in the field-redefinition \eqref{eq:FieldRedefinition}.
Hence, this solution represents homogeneous $\phi$-configurations as we explicitly discussed in \cite{Schmidt2024}.
The conserved momentum that is conjugate to this homogeneous field is an instance of Weinberg's conservation of adiabatic modes in cosmology \cite{Weinberg2003}.
From the quantum-mechanical perspective, the conformal factor $a^w(\eta)$ describes a resonance that occurs at the boundary between the scattering continuum and the bound states.
As a further consequence, the scattering potential satisfies the Riccati-equation 
\begin{equation}
    V(\eta) = - W'(\eta) + W^2(\eta), \quad W(\eta) = - w \frac{a'(\eta)}{a(\eta)},
    \label{eq:RiccatiForm}
\end{equation}
where $a'/a$ is the conformal Hubble rate. 
In the presence of a field mass or non-minimal coupling, there are correction terms to \cref{eq:RiccatiForm}.
The requirement of asymptotic staticity leads to the property $\lim_{\eta \to \pm \infty} W(\eta) = 0$, which will become important in the following.
Zero-energy resonances are peculiar in the context of quantum scattering for multiple reasons:
\begin{enumerate}
    \item[(i)] They are not bound-states since they are not $L^2$-normalizable and thus also do not contribute to a spectral resolution of the identity \cite{Chadan1989}. 
    These are exactly the kind of states Jost would claim to be `false zeros' of his Jost function \cite{Jost1947}. 
    He deliberately considered the upper half plane in complexified momentum space and excluded the real line.
    \item[(ii)] They count as half a bound-state in Levinson's theorem \cite{Wellner1964,Boya2007}, which counts the number of bound-states based on the phase difference between asymptotic scattering states \cite{Boya2008}.
    \item[(iii)] Whereas in the generic case, the zero-mode is totally reflected, $r_\pm(0) = -1$ and $t(0) = 0$, the existence of a zero-energy resonance entails that $t(0) = \sin \alpha$ and $r_\pm(0) = \pm \cos \alpha$ for some phase $\alpha$ \cite{Chadan1989}. 
\end{enumerate}
Hence, the mathematical literature classifies the existence of a zero-energy solution as an `exceptional point' (a thorough discussion can be found in \cite[section XVII.1.4]{Chadan1989}).
Since this exceptional case corresponds to the common case for cosmological particle production, it is interesting to investigate it in further detail.
While we did already discuss implications of point (iii) onto the cosmological power spectrum \cite{Schmidt2024}, we want to investigate the zero-energy resonance in further detail here, with a particular emphasis on its connection to (analog) cosmic singularities and its role in the inverse problem.
For simplicity, we restrict to the spatially flat case, $\kappa_\mathrm{sp} = 0$, for the remainder of this work.

\subsubsection{Bound States and Cosmic Singularities}
\begin{figure*}
    \includegraphics[width = \textwidth]{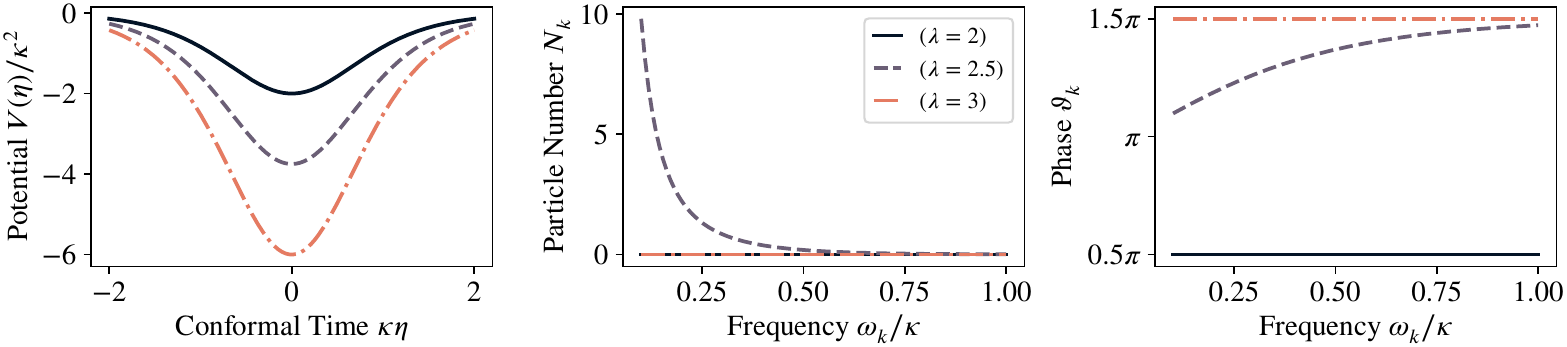}
    \caption{Left: visualization of the Pöschl-Teller potential hole. Particle number $N_k$ (center) and phase $\vartheta_k$ (right) of the spectrum $S_k$.}
    \label{fig:PöschlTellerSpectrum}
\end{figure*}
\begin{figure*}
    \includegraphics[width=\textwidth]{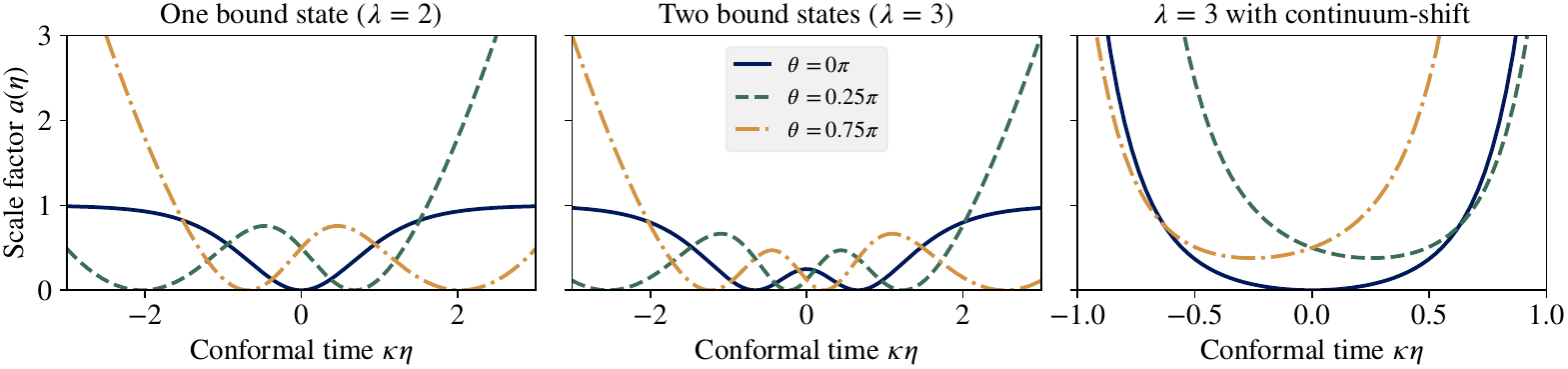}
    \caption{Cosmological evolutions that are coresponding to the Pöschl-Teller hole in $D=2$ spatial dimensions (left, $\lambda = 2$, and center, $\lambda = 3$) 
    of the form $a(\eta) = [c_1 P_{\lambda-1}(\tanh \kappa \eta) + c_2 Q_{\lambda-1}(\tanh \kappa \eta) ]^2$, where $c_1 = \cos \theta$, $c_2 = \sin \theta$ and $P_{\lambda-1},Q_{\lambda-1}$ are Legendre functions of first and second kind, respectively.
    A continuum shift (right panel) lets the zero-energy solution be the ground state, thereby removing the singularity.} 
    \label{fig:PöschlScalFig}
\end{figure*}
The central observation is that zeros of the scale-factor $a$ (big-crunch--big-bang singularities) correspond to nodes of the zero-energy solution $a^w(\eta)$.
By Sturm's oscillation theorem \cite{Teschl2012} (see also \cite[Thm.~2.3]{DeiftTrubowitz1979} for the zero-energy case on the line and \cite{Moriconi2007} for an intuitive argumentation), each of these nodes is associated to a bound-state.
Notice however that the particle production process is agnostic to such cosmic singularities by virtue of the scattering analogy, since potentials that possess bound-states can reflect waves without any irregularities 
-- a phenomenon known as quantum reflection \footnote{We are here assuming that the standard prerequisites for quantum field theory in curved spacetime still hold without any quantum gravity modifications. This is questionable for a real big-crunch--big-bang singularity.}.

Criteria under which bound-states, and thus analog cosmological singularities occur, have been derived in the literature.
As shown in \cite{Buell1995}, any potential that satisfies
\begin{equation}
    \int_{-\infty}^{\infty} \mathrm{d}\eta \, V(\eta) < 0
    \label{eq:BoundStateCondition}
\end{equation}
possesses at least one bound-state. 
Now, in the cosmological setting we have that
\begin{equation}
    \int_{-\infty}^{\infty} \mathrm{d}\eta \, V(\eta) = w^2 \int_{-\infty}^{\infty} \, \mathrm{d}\eta \, \left(\frac{a'(\eta)}{a(\eta)}\right)^2 > 0
    \label{eq:BoundStateConditionCosmo}
\end{equation}
by virtue of the Riccati-equation \eqref{eq:RiccatiForm} and asymptotic staticity, $\lim_{\eta \to \pm \infty} W(\eta) = 0$.
However, \cref{eq:BoundStateConditionCosmo} does not indicate that potentials of the form \eqref{eq:ScatteringPot} cannot have bound-states,
since the criterion \eqref{eq:BoundStateCondition} is sufficient but not necessary (a counter-example involving a $\delta$-potential is discussed in \cite{Buell1995}).

\subsubsection{Example: Pöschl-Teller Potential}

The preceding discussion can be best exemplified with the Pöschl-Teller potential \cite{Poeschl1933,Fluegge1999}, since it is entirely determined by its bound states.
It is defined by  
\begin{equation}
    V(\eta) = - \kappa^2 \lambda (\lambda-1) \sech^2(\kappa \eta)
    \label{eq:PöschlTellerPotential} 
\end{equation}
and has $\lceil \lambda \rceil - 1$ bound-states. 
For integer $\lambda = N+1$ this is precisely the reflectionless landscape with $N$
bound states that, taken as initial data for the Korteweg--de Vries equation, separates
into $N$ solitons \cite{Gardner1967, Dunajski2010}.
Indeed, calculating the reflection coefficient and the phase-shift, one finds 
\begin{equation}
    \abs{r_k}^2 = \cos^2(\varphi_e - \varphi_0), \quad \varphi_\mathrm{e} - \varphi_0 = \arctan \left[ \frac{\sinh(\pi k / \kappa)}{\sin (\pi \lambda)}\right],
    \label{eq:PöschlSpectrum}
\end{equation}
where $\varphi_\mathrm{e} - \varphi_0$ is the accumulated phase \cite{Fluegge1999}, which is equal for all modes if $\lambda$ is integer.
In \cref{fig:PöschlTellerSpectrum}, this potential together with its analog cosmological scattering data is shown, showcasing the occurrence of a quantum recurrence across the whole spectrum if $\lambda$ is integer.
The cosmological evolution corresponding to the Pöschl-Teller potential can be found using the zero-energy resonance, and is displayed in \cref{fig:PöschlScalFig} for the case $D=2$. 
As shown in the left and central panel, increasing $\lambda$ by unity adds another bound-state to the Pöschl-Teller landscape and hence leads to an additional singularity of the scale-factor.
A continuum shift of the potential landscape by its ground-state energy cures the cosmological singularities (see right panel).
Despite these singularities, the scattering problem is always well-defined and integrable. 

\subsection{Expansion-Contraction Duality}
\label{subsec:Duality}
In a seminal paper, Wands \cite{Wands1999} showed that a certain contracting spacetime also produces a scale-invariant power spectrum, like an exponential expansion, based on both scenarios leading to the same effective mass.
Here, the case shall be made that such isospectralities may also be achieved through different effective masses (potentials in the scattering analogy). 

In scattering theory, it is well-known that isospectral potentials -- in the sense that their spectra only differ by a phase -- can be obtained systematically via Darboux-transformations \cite{Chadan1989}.
For instance, the isospectral partner $\tilde V(\eta)$ of a potential $V(\eta)$ that can be written in the Riccati-form \eqref{eq:RiccatiForm} is $\tilde V(\eta) = W'(\eta) + W^2(\eta)$
in the sense that their scattering amplitudes satisfy 
\begin{equation}
    \tilde r(k) = - r(k), \quad \tilde t(k) = t(k),
    \label{eq:IsospectralAmplitudes}
\end{equation}
as we explicitly show in \cref{sec:Isospectrality} under usage of the asymptotics $\lim_{\eta \to \pm \infty} W(\eta) = 0$.
\Cref{eq:IsospectralAmplitudes} is the result of one of the Darboux-transformations discussed in \cite[section XVII.3.2]{Chadan1989}.
This entails that the particle number and the amplitude in \cref{eq:ParticleNumber,eq:Amplitude} are equivalent, but the phase differs by a $\pi$-shift, rendering it a robust discriminator among models.
In particular, if $V(\eta)$ has a zero-energy resonance $a^w(\eta)$, one finds that $\tilde V(\eta)$ also has a zero-energy resonance given by the inverse $a^{-w}(\eta)$.
As a consequence, the scale-factor $\tilde a(\eta)$ that generates $\tilde V(\eta)$ via the same form as in \cref{eq:ScatteringPot} follows from the differential equation
\begin{equation}
    \begin{aligned}
    W'(\eta) + W^2(\eta) 
    &= \frac{D-1}{2} \left[\frac{D-3}{2} \left( \frac{\tilde a'(\eta)}{\tilde a(\eta)} \right)^2 + \frac{\tilde a''(\eta)}{\tilde a(\eta)} \right] \\
    &=  \frac{D-1}{2} \left[ \frac{D+1}{2} \left( \frac{a'(\eta)}{a(\eta)}\right)^2 - \frac{a''(\eta)}{a(\eta)} \right]
    \end{aligned}
\end{equation}
which is solved by $\tilde a(\eta) = 1 / a(\eta)$.
Consequently, for each expanding scenario $a(\eta)$ there exists a dual contracting scenario $\tilde a(\eta) = 1 / a(\eta)$
which leads to the same particle number and amplitude but differs by a $\pi$-phase in the power spectrum.
This duality map is known as the special limit transformation in the inverse-scattering literature \cite{Degasperis1987}.

\section{Inverse Quantum Scattering Theory in One Dimension}
\label{sec:InverseScattering}
The purpose of this section is to familiarize the interested reader with the inverse quantum scattering in one dimension.
Our summary mainly follows \cite{Faddeev1976,Newton1980,Chadan1989}, is not extensive and does not cover the mathematical depth and rigor required to derive the presented results.
Further basic information is collected in \cref{app:Scat_Bound}.
For an extensive discussion, the reader is referred to \cite{Chadan1989} and references therein. 

\subsection{The Direct Problem}
It is common to classify one-dimensional scattering potentials $V(\eta)$ into spaces $L_1^p$, where $V(\eta)$ needs to be real, locally integrable and is an element of $L_1^p$ iff
\begin{equation}
   \int_{-\infty}^{\infty} \mathrm{d}\eta \, (1 + \lvert \eta \rvert^p) \, \lvert V(\eta) \rvert < \infty.
\end{equation}
Inverse scattering usually examines the Faddeev class $L_1^1$, for which a solution to the one-dimensional Schrödinger equation needs to be continuously differentiable
and only finitely many bound states exist.
The formalism can be generalized to less regular potentials, like the Dirac-delta potential \cite{Chadan1989}.

The two kinds of scattering solutions have the asymptotic form 
\begin{equation}
    \psi_-(k,\eta) = \begin{cases}
        \mathrm{e}^{\mathrm{i}k\eta} + r_-(k) \, \mathrm{e}^{-\mathrm{i} k \eta}, \quad \eta \to - \infty \\
        t(k) \, \mathrm{e}^{\mathrm{i} k \eta}, \quad \eta \to \infty
    \end{cases}
    \label{eq:ScatteringSolution1}
\end{equation}
or 
\begin{equation}
    \psi_+(k,\eta) = \begin{cases}
        t(k) \, \mathrm{e}^{-\mathrm{i} k \eta} \quad \eta \to - \infty, \\
        \mathrm{e}^{-\mathrm{i}k\eta} + r_+(k) \, \mathrm{e}^{\mathrm{i} k \eta},\quad \eta \to \infty,
    \end{cases}
    \label{eq:ScatteringSolution2}
\end{equation}
where the scattering amplitudes are collected in the S-matrix 
\begin{equation}
    S(k) = \begin{pmatrix}
        t(k) & r_+(k) \\
        r_-(k) & t(k)
    \end{pmatrix},
\end{equation}
with $r_\pm(k)$ as the amplitude of a wave that is reflected in the positive (negative) direction $\eta \to \pm \infty$. 

Faddeev-Marchenko theory centers around different kinds of solutions to the Schrödinger equation, so-called Jost solutions, which are defined by the asymptotic behavior $\lim_{\eta \to \pm \infty}  \mathrm{e}^{\mp \mathrm{i} k \eta}f_\pm(k,\eta) = 1$,
and related to the scattering solutions via $\psi_\pm(k,\eta) = t(k) f_\mp(k,\eta)$.
The Jost solutions satisfy the familiar Volterra integral equation 
\begin{equation}
    f_\pm(k,\eta) = \exp(\pm \mathrm{i} k \eta) + \int_{-\infty}^\infty G_\mp(\eta - \eta',k)\, V(\eta')\, f_\pm(k,\eta') \, \mathrm{d}\eta',
    \label{eq:VolterraJost}
\end{equation}
where $G_\mp(\eta,k) = \mp \Theta(\mp\eta) \sin(k \eta)/k$ is the advanced (-) and retarded (+) Green's function of $\left(\partial_\eta^2 + k^2\right)$ in terms of the Heaviside-function $\Theta$.
Resumming the iterated series of \eqref{eq:VolterraJost} yields the exact Levin-representation of the Jost solutions \cite{Levin1956}
\begin{equation}
    f_\pm (k,\eta) = \exp(\pm \mathrm{i} k \eta) + \int_{-\infty}^\infty K_\pm(\eta,\eta')  \exp(\pm \mathrm{i} k \eta') \mathrm{d}\eta',
    \label{eq:LevinRepJost}
\end{equation}
which can be understood as a linear transformation of the free solution onto the Jost solutions.
The transformation kernel $K_\pm(\eta,\eta')$ has triangular support on $\pm(\eta' - \eta) > 0$.
Its importance in inverse scattering theory stems from the relation of its diagonal to the scattering potential \cite{AgranovichMarchenko2020}
\begin{equation}
    K_+(\eta,\eta) = \frac{1}{2} \int_\eta^{\infty} \mathrm{d}s \, V(s), \quad K_-(\eta,\eta) =  \frac{1}{2} \int_{-\infty}^\eta \mathrm{d}s \, V(s),
    \label{eq:TransformationKernelPotential}
\end{equation}
such that the potential can be obtained via
\begin{equation}
    V(\eta) = - 2 K_+'(\eta,\eta) = 2 K_-'(\eta,\eta).
    \label{eq:PotentialDifferentiation}
\end{equation}

\subsection{The Inverse Problem}
\subsubsection{Faddeev-Marchenko Equations}
The inverse problem in one dimension is stated via the Faddeev-Marchenko equations (FMEs)
\begin{widetext}
    \begin{align}
    K_+(\eta,\eta') + M_+(\eta + \eta') + \int_\eta^{\infty} \dd \eta'' \, K_+(\eta,\eta'') M_+(\eta'' + \eta') = 0 \quad (\eta' > \eta), \label{eq:FaddeevMarchenkoUpper} \\
    K_-(\eta,\eta') + M_-(\eta + \eta') + \int_{-\infty}^\eta \dd \eta'' \, K_-(\eta,\eta'') M_-(\eta'' + \eta') = 0 \quad (\eta' < \eta), \label{eq:FaddeevMarchenkoLower}
\end{align}
\end{widetext}
see \cref{sec:FM_Derivation} for their derivation.
Therein, the spectral data is contained in the Marchenko functions 
\begin{equation}
        M_\pm(\eta) =  \int_{-\infty}^\infty \frac{\dd k }{2\pi}\, \mathrm{e}^{\pm \mathrm{i} k \eta} \, r_\pm(k) + \sum_{n=0}^{N-1} (C_n^\pm)^{-2} \mathrm{e}^{\mp \kappa_n \eta},
        \label{eq:MarchenkoFunction}
\end{equation}
with the bound-state weight
\begin{equation}
  (C_n^\pm)^{-2} = \left[ \int_{-\infty}^{\infty} \mathrm{d}\eta \, f_\pm(\mathrm{i}\kappa, \eta)^2 \right]^{-1}
\end{equation}
The FMEs \eqref{eq:FaddeevMarchenkoLower} and \eqref{eq:FaddeevMarchenkoUpper} are Fredholm integral equations for the transformation kernels $K_\pm(\eta,\eta')$ in the second variable $\eta'$ at fixed $\eta$.
Each of the FMEs represents information corresponding to one of the half-strips of the real line. 
This structure is inherited from the radially symmetric three-dimensional case where the boundaries are at the origin and asymptotic infinity.
In one dimension, there is one origin and two notions of infinity, hence two equations.
For parity-symmetric potentials (such as the benchmarks in \cref{subsec:NumMethods}), 
the positive and negative half-strip Jost solutions are related by the $\mathbb{Z}_2$-inversion, $f_-(k,\eta) = f_+(k,-\eta)$ 
which entails for the transformation kernels that $K_-(\eta, \eta') = K_+(- \eta, -\eta')$.

\subsubsection{Uniqueness}
Potentials in the Faddeev class $L_1^1$ are uniquely determined by their reflection amplitude as well as their bound-state energies and normalization constants \cite{Faddeev1964,DeiftTrubowitz1979}.
This includes the exceptional case of zero-energy resonances.
Degasperis \& Sabatier \cite{Degasperis1987} classified potentials beyond the Faddeev class according to the asymptotics of their zero-energy solutions.
Cosmological potentials which are generated from an asymptotically static spacetime (cf.~\cref{sec:CosmoScattering}) fall into the unambiguous class containing potentials that are uniquely fixed by the scattering- and bound-state data.
Here one should clarify that the expansion-contraction duality discussed in \cref{subsec:Duality} is not an ambiguity in their sense, as the reflection amplitudes in \cref{eq:IsospectralAmplitudes} genuinely differ.

\subsection{Exact Solutions}
\label{subsec:ExactSolutions}
While this inverse problem is notoriously difficult to solve analytically, it admits closed form solutions if the scattering potential is \emph{unimodal} \cite{Boya2008}, i.e.\@ its scattering data can be described in terms of a single phase.
For example, the Pöschl-Teller potential \eqref{eq:PöschlTellerPotential} is unimodal with phase $\varphi_e - \varphi_0$ from \cref{eq:PöschlSpectrum}.
Other exactly solvable situations include the $\delta$- and $\delta'$-potential.
The $\delta$-pair potential on the other hand is not unimodal, since the peak separation induces another phase to the problem \cite{Boya2008}.

\subsubsection{Delta-Potential}
\label{subsubsec:Delta_Sol}
For potentials with compact support (such as those occurring in a quantum simulation of cosmological particle production), a few useful identities involving Fourier-transforms of the scattering amplitudes exist (cf.\@ \cref{subsec:PotentialCompact_Sol}) and benefit a solution to the inverse problem.
As an example consider the $\delta$-potential $V(\eta) = 2 \alpha \delta(\eta)$ with $\alpha \in \mathbb{R}$,
which is a well-known textbook problem resulting in the scattering amplitudes \cite{Fluegge1999}
\begin{equation}
    r_\pm(k) = - \mathrm{i} \alpha (k + \mathrm{i} \alpha)^{-1}, 
    \quad t(k) = k (k + \mathrm{i}\alpha)^{-1}.
\end{equation}
The transmission coefficient has a pole at $k = - \mathrm{i} \alpha$ which lies in the physical sheet, $\mathrm{Im} \, k > 0$, for $\alpha <0$, and in that case hence represents a genuine bound state with normalization $(C^\pm)^{-2} = \lvert \alpha \rvert$.
For the Fourier-transforms of the reflection amplitudes we find
\begin{equation}
        \int_{-\infty}^\infty \frac{\mathrm{d}k}{2\pi}  \,  r_\pm(k) \, \mathrm{e}^{\pm \mathrm{i} k \eta} 
        =  - \lvert \alpha \rvert  \exp(- \lvert \alpha \eta \rvert) \, \Theta \left(\mp \mathrm{sgn}(\alpha)\, \eta\right),
\end{equation}
and the Marchenko functions are 
\begin{equation}
    \begin{aligned}
        M_\pm(\eta) &= - \alpha \, \mathrm{e}^{- \alpha \lvert \eta \rvert} \, \Theta(\mp \eta),
    \end{aligned}
    \label{eq:MarchenkoDelta}
\end{equation}
which by virtue of the FMEs entails that $V(\eta) = 0 $ for $\eta > 0$ and $\eta < 0$ for both positive and negative values of $\alpha$.
Due to this behavior, we can calculate the transformation kernels directly through \cref{eq:KernelPositive,eq:KernelNegative} with $r_\pm(k) / t(k) = - \mathrm{i}\alpha /k$ to find 
\begin{equation}
    \begin{aligned}
        K_+(\eta,\eta') 
        &= \alpha \, \Theta (- \eta - \eta') \label{eq:KernelPositiveDelta}
    \end{aligned}
\end{equation}
for $\eta' > \eta$ with $\eta < 0$, and similarly
\begin{equation}
    K_-(\eta,\eta') = \alpha \, \Theta(\eta + \eta')
    \label{eq:KernelNegativeDelta}
\end{equation}
for $\eta' < \eta$ and $\eta > 0$.
Since
\begin{equation}
    V(\eta) = - 2 K_+'(\eta, \eta) = 2\alpha \delta (\eta),
\end{equation}
for $\eta > 0$,
and 
\begin{equation}
    V(\eta) = 2 K_-'(\eta, \eta) = 2\alpha \delta (\eta),
\end{equation}
for $\eta < 0$, agree, we can conclude that $V(\eta) = 2 \alpha \delta(\eta)$ on the whole line.
Considering the support of the Marchenko functions, it is possible to show that $K_+(\eta, \eta') = 0$ for $\eta' > - \eta$ as a consequence of the FMEs, 
such that together with $\eta' > \eta$ the kernel \eqref{eq:KernelPositiveDelta} can only be nonzero for $\eta < 0$.
A similar result holds for $K_-$ in context of \cref{eq:KernelNegativeDelta}, so we have the stronger result 
\begin{equation}
    \begin{aligned}
        K_+(\eta,\eta') &= \alpha \, \Theta(- \eta - \eta') \quad &(\eta' > \eta) \\
        K_-(\eta,\eta') &= \alpha \, \Theta(\eta + \eta') \quad &(\eta' < \eta).
    \end{aligned}
    \label{eq:TransformationKernelDeltaFinal}
\end{equation}

\subsubsection{Pöschl-Teller Potential}
\label{subsubsec:Pöschl_Sol}
Consider again the Pöschl-Teller potential \eqref{eq:PöschlTellerPotential} with a single bound state $(\lambda = 2)$.
Since the scattering continuum is absent, the FMEs reduce to an algebraic set of linear equations which can be solved by a separable (also called degenerate) transformation kernel.
In the case of a single bound state, the transformation kernel is \cite{Chadan1989}
\begin{align}
    K_+(\eta, \eta') = - 2 \kappa \frac{\exp (- \kappa (\eta + \eta')) }{1 + \exp(-2 \kappa \eta)} \quad (\eta' > \eta), \\
    K_-(\eta, \eta') = - 2 \kappa \frac{\exp (\kappa (\eta + \eta')) }{1 + \exp(2 \kappa \eta)} \quad (\eta' < \eta),
\end{align}
as can be derived from the Marchenko functions $M_\pm(\eta) = 2 \kappa \exp(\mp \kappa \eta)$.
The calculation follows along the lines of Sabatier's algorithm for rational reflection amplitudes (cf.\@ \cref{subsec:RationalScat_Sol}),
for which the Fourier-transforms that appear in the Marchenko functions have the same form as the bound-state terms.

\subsection{Numerical Method}
\label{subsec:NumMethods}
\begin{figure*}
    \centering
    \resizebox{\textwidth}{!}{%
    \begin{tikzpicture}[
        node distance = 12mm,
        every node/.style = {font=\small},
        box/.style = {draw, thick, rounded corners, align=center, minimum height=10mm, inner sep=6pt},
        proc/.style = {midway, above=2.5mm, font=\small\itshape},
        >={Stealth[length=3mm]}
    ]
        \node[box] (a)   {$a(\eta)$};
        \node[box, right=18mm of a] (V) {$V(\eta)$};
        \node[box, right=18mm of V] (r) {$r_\pm(k)$};
        \node[box, right=18mm of r] (M) {$M_\pm(\eta)$};
        \node[box, right=18mm of M] (K) {$K_\pm(\eta,\eta')$};
        \node[box, right=18mm of K] (Va) {$a(\eta)$};

        \draw[->] (a) -- (V) node[proc] {Differentiate};
        \draw[->] (V) -- (r) node[proc] {Numerov};
        \draw[->] (r) -- (M) node[proc] {FFT};
        \draw[->] (M) -- (K) node[proc] {Nyström};
        \draw[->] (K) -- (Va) node[proc] {Integrate};
    \end{tikzpicture}%
    }
    \caption{The reconstruction scheme. 
    A scale-factor $a(\eta)$ fixes the potential $V(\eta)$, whose reflection amplitude $r_\pm(k)$ is obtained from a Numerov-solver; their Fourier transform gives the Marchenko functions $M_\pm(s)$.
    These enter the FME which is solved via a Nyström-scheme for the transformation kernel $K_\pm$, from which $a$ is recovered via the zero-energy Jost solution $f_\pm(0,\eta)$.}
    \label{fig:Pipeline}
\end{figure*}
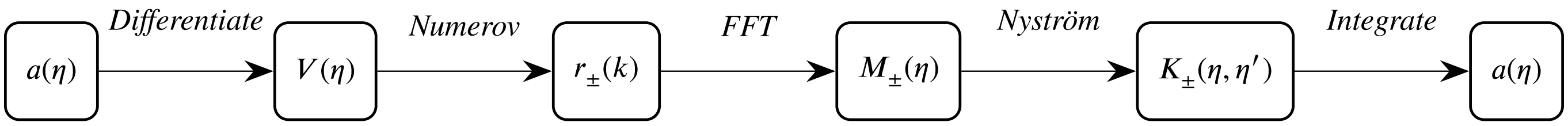

For the numerical analysis we restrict to two spatial dimensions, $D = 2$.
Our numerical reconstruction scheme is visualized in \cref{fig:Pipeline}.
We feed in a model scale-factor, compute its potential via \cref{eq:ScatteringPot} and solve the scattering problem via the Numerov-algorithm \cite{NumericalRecipes2007}.
The Marchenko functions \eqref{eq:MarchenkoFunction} are computed via the Fast-Fourier-Transform (FFT) and its inverse; respectively applied to the numerically obtained $r_\pm(k)$.
The FMEs are then solved for the transformation kernels $K_\pm(\eta,\eta')$ via the Nyström-method (described in the following subsection).
Each of the transformation kernels yields the scale factor of interest via
\begin{equation}
    a(\eta) = a_\mathrm{f} \, f_+^2(0,\eta), \quad a(\eta) = a_\mathrm{i} \, f_-^2(0,\eta),
    \label{eq:ScaleFactorReconstruction}
\end{equation}
with the zero-energy Jost solutions (cf.\@ \cref{eq:LevinRepJost})
\begin{equation}
\begin{aligned}
    f_+(0,\eta) &= 1 + \int_\eta^\infty \mathrm{d} \eta' K_+(\eta,\eta'), \\
    f_-(0,\eta) &= 1 + \int_{-\infty}^\eta \mathrm{d} \eta' K_-(\eta,\eta').
\end{aligned}
\label{eq:ZeroEnergyJost}
\end{equation}
Note that \cref{eq:ZeroEnergyJost} is more numerically stable than \cref{eq:PotentialDifferentiation} due to an integration (rather than differentiation) of the transformation kernel.
Hence, solving for the zero-energy solutions rather than the potential (the medium) is the traditional route to solve the inverse problem in seismics and acoustics \cite{Kay1972,BerrymanGreene1980,Coen1981}, as we have mentioned initially.
This setting has a direct precursor in the inverse acoustic problem treated by Deift \& Trubowitz \cite{DeiftTrubowitz1979}
with the zero-energy Jost solution as the inverse square root of the sound speed.

\subsubsection{Nyström-Solver}
The FMEs \eqref{eq:FaddeevMarchenkoUpper} and \eqref{eq:FaddeevMarchenkoLower} are Fredholm equations of the second kind in $\eta'$ at fixed $\eta$.
It is standard to solve these via the Nyström method \cite{NumericalRecipes2007}, which we summarize for $K_+$ (similarly for $K_-$).
We place $n$ nodes $\eta''_1 < \dots < \eta''_n$ on the truncated integration interval and collocate
\cref{eq:FaddeevMarchenkoUpper} at these nodes, which results in the linear system
\begin{equation}
    \sum_{j=1}^{n} \Big[ \delta_{ij} + W(\eta''_i, \eta''_j) \Big] \, K_+(\eta,\eta''_j)
    = - M_+(\eta + \eta''_i),
    \label{eq:NyströmSystem}
\end{equation}
for the $n$ unknown kernel values $K_+(\eta,\eta''_i)$ where $i = 1,\dots,n$. 
These in turn serve as nodes to infer the kernel at an arbitrary output point via a final quadrature,
\begin{equation}
    K_+(\eta,\eta') = - M_+(\eta + \eta')
    - \sum_{j=1}^{n} W(\eta',\eta''_j) \, K_+(\eta,\eta''_j),
    \label{eq:NyströmEvaluation}
\end{equation}
where the quadrature weights
\begin{equation}
    W(\eta',\eta''_j) =
    \begin{cases}
        w_j \, M_+(\eta''_j + \eta'), & \mathrm{(GL)}\\[4pt]
        \displaystyle\int \dd \eta'' \, \Lambda_j(\eta'') \, M_+(\eta'' + \eta'), & \mathrm{(PWL)}
    \end{cases}
    \label{eq:NyströmWeights}
\end{equation}
are chosen according to the smoothness of $M_+$. 
The first line is an ordinary Gauss-Legendre rule (GL) with nodes $\eta''_j$ and weights $w_j$, which is appropriate when $M_+$ is analytic.
The second line is a piecewise-linear product integration (PWL), which applies for non-analytic $M_+$. 
Therein $\Lambda_j$ is the hat function centered on $\eta''_j$. 

Let us give two specific examples where both rules find application:
For the cosmological models that are discussed in \cref{sec:NumericalReconstruction}, the reflection amplitude $r_+(k)$ decays exponentially such that $M_+$ is analytic, and the Gauss-Legendre rule with $n = 80$ nodes per row converges.
On the other hand, for the $\delta$-potential benchmark, the reflection amplitude $r_+(k)$ follows the power-law $r_+(k) \propto k^{-1}$, and the Marchenko function $M_\pm$ carries a step discontinuity
for which a global polynomial quadrature rule fails. 
We then use a PWL with $n = 200$ hat functions and evaluate the integrals in \cref{eq:NyströmWeights} by a $6$-point Gauss rule on sub-cells split exactly at the
discontinuity, such that no quadrature cell overlaps a jump.

\subsubsection{Choosing the Grid}

The Marchenko functions are known on an \emph{external} strip $\eta \in [-L_\mathrm{ext}, L_\mathrm{ext}]$, fixed by the
Fourier transform of $r_\pm(k)$, whereas the transformation kernels are returned
on a \emph{region of interest} (ROI) $[-L, L] \times [-L,L]$ strictly inside the grid spanned by the external strip, $L < L_\mathrm{ext}$. 
This focus on an ROI is demanded by \cref{eq:FaddeevMarchenkoUpper}: Computing
$K_+(\eta,\cdot)$ at a given $\eta$ requires knowledge of $M_+$ at arguments $\eta'' + \eta'$.
Hence, when both kernel arguments lie in the ROI their sum ranges over $[-2L,2L]$, so $L_\mathrm{ext} \geq 2L$
is required. We choose the zoom factor $L_\mathrm{ext} / L = 2$. 
The parameter $L$ itself is fixed by the convergence of the integral \eqref{eq:ZeroEnergyJost}.
Depending on how fast the potential decays into the positive (negative) $\eta$-direction, $L$ needs to be chosen sufficiently large to ensure that the integration of $f_+$ (respectively $f_-$) does not run out of the ROI.
For the cosmological models discussed in \cref{sec:NumericalReconstruction}, this integral converges satisfactorily for
$L = 24\,\eta_0$, where $\eta_0$ is a characteristic scale of the cosmological system (to be discussed in \cref{sec:NumericalReconstruction}).
We sample the external strip at $N = 1201$ points of which the $601$ innermost span the ROI.

\subsubsection{Benchmarks}
\begin{figure}
    \includegraphics[width=\columnwidth]{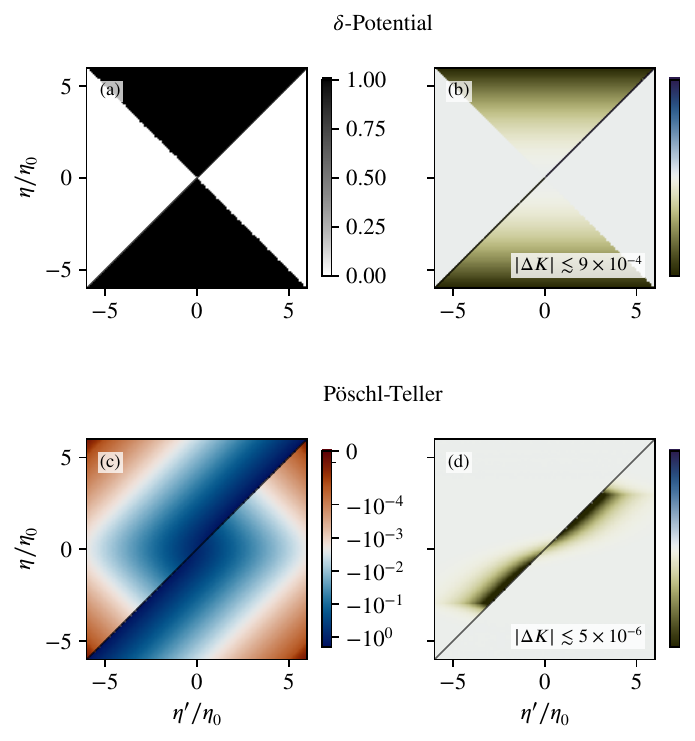}
    \caption{Transformation kernel $K(\eta,\eta')$ for the two exact benchmarks of \cref{subsubsec:Delta_Sol,subsubsec:Pöschl_Sol} (upper row: $\delta$-potential, lower row: Pöschl-Teller potential).
    Here we introduced the characteristic scale $\eta_0 = \lvert \alpha \rvert^{-1}$ (for the $\delta$-potential) and $\eta_0 = \kappa^{-1}$ (for the Pöschl-Teller potential).
    The numerical solution (left column) relatively deviates from the closed form as shown in the right column where we plot $\Delta K = (K_\mathrm{num}-K_\mathrm{exact})/\max|K_\mathrm{exact}|$ with a colorbar centered around zero
    and an order-of-magnitude-bound given in the lower-right of each panel.}
    \label{fig:BenchmarkKernels}
\end{figure}

We benchmark the Numerov-FFT-Nyström pipeline (cf.\@ \cref{fig:Pipeline}) via the analytically solvable reference models discussed in \cref{subsec:ExactSolutions}.
The $\delta$-potential is a reference case for the scattering part of the Marchenko function which can feature a sharp jump in $\eta$ due to analyticity properties of $r_\pm(k)$ (see for instance \cref{eq:MarchenkoDelta}).
On the other hand, the bound-state part in $M_\pm(\eta)$ is smooth and benchmarked with the reflectionless Pöschl-Teller scenario.
\Cref{fig:BenchmarkKernels} displays the numerically solved kernel alongside its relative deviation from the closed form, showcasing agreement at the sub-permille level.

\section{Application to Cosmological Particle Production: Logistic Expansions}
\label{sec:NumericalReconstruction}

Restricting to the non-singular case where the scale-factor has no zeros, we only need to consider the scattering part of the Marchenko function (cf.\@ \cref{eq:MarchenkoFunction}).
As the zero-energy resonance is not $L^2$-normalizable, it does not contribute to a spectral resolution of the identity \cite{Chadan1989} and consequently does not enter $M_\pm(\eta)$.

\subsection{The Target Models}
\label{subsec:TargetModels}
\begin{figure}
    \includegraphics[width=\columnwidth]{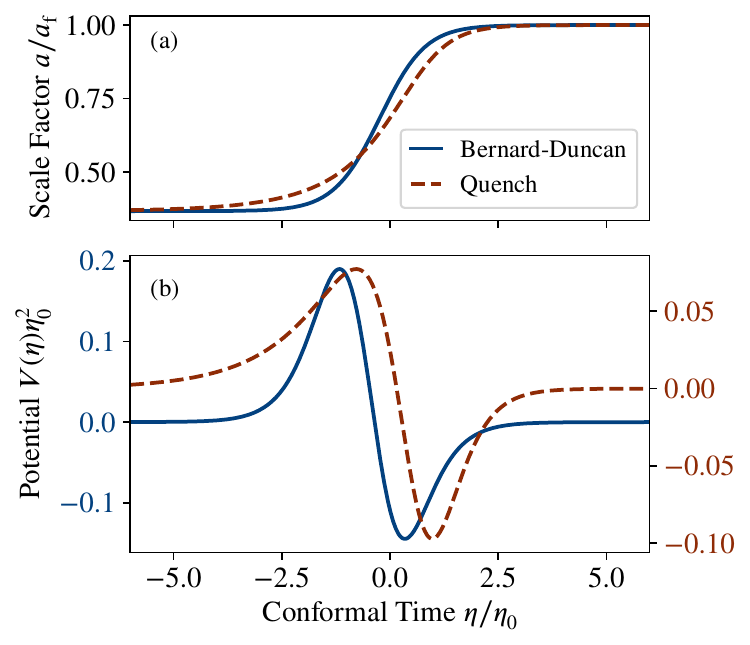}
    \caption{Top: The scale factor $a(\eta)/a_\mathrm{f}$ of both models. Bottom: The associated potential $V(\eta) \, \eta_0^2$ for $a_\mathrm{f} / a_\mathrm{i} = \mathrm{e}$, where we use a twin vertical axis for the quench model since its potential is a factor $\sim 2.5$ shallower.
    To contextualize with intuition from scattering theory, we note that the Bernard-Duncan potential can be fitted to a smoothed linear combination of $\delta$- and $\delta'$-potentials.}
    \label{fig:Models}
\end{figure}

\begin{figure}
    \includegraphics[width=\columnwidth]{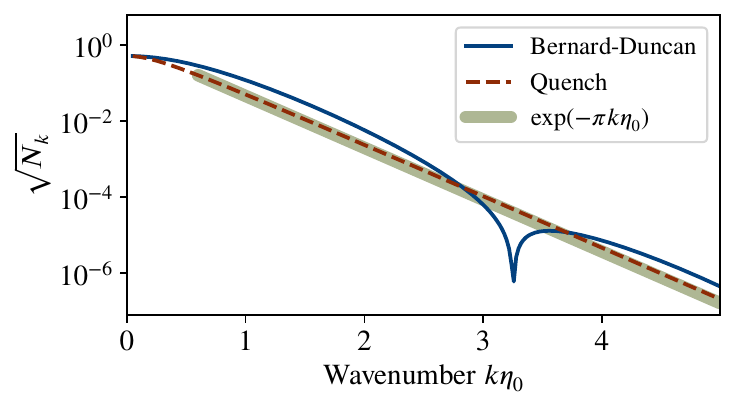}
    \caption{Exponential decay of the particle production signal for the Bernard-Duncan and the quench model. 
    The decay rate can be calculated with the Paley-Wiener theorem (see main text).
    The dip of the Bernard-Duncan signal corresponds to both poles of the potential with the same imaginary part competing as a consequence of unequal real parts.
    The quench model decays monotonically as it does not have such a coincident pair of poles, since the complex zero of $a(t)$ is mapped to infinity in conformal time.}
    \label{fig:ReflectionDecay}
\end{figure}

In the following we consider two classic examples with identical asymptotics $a_\mathrm{i} = \mathrm{e}^{-1}$, $a_\mathrm{f} = 1$ but slightly different histories:
On the one hand, we have the \emph{Bernard-Duncan} model \cite{BernardDuncan1977}
\begin{equation}
    a^2(\eta) = \frac{1}{2} \left(a_\mathrm{f}^2 + a_\mathrm{i}^2 + (a_\mathrm{f}^2 - a_\mathrm{i}^2 ) \tanh(\eta/\eta_0) \right),
    \label{eq:BernardDuncanModel}
\end{equation}
which resembles logistic growth and is an archetypal textbook-example for quantum field theory in curved spacetime known from Birrell \& Davies \cite{BirrellDavies1982}, 
and strictly speaking centered around a massive field in a conformal vacuum \cite{Ford2021}.
We will nevertheless investigate the Bernard-Duncan scalefactor outside its original setting.
From an operational perspective, it seems more natural to consider a logistic growth of $a(t)$ rather than $a^2(\eta)$, i.e.\@ 
\begin{equation}
    a(t) = \frac{1}{2} \left(a_\mathrm{f} + a_\mathrm{i} + (a_\mathrm{f} - a_\mathrm{i}) \tanh(t/t_0) \right),
    \label{eq:quench}
\end{equation}
which is known as a \emph{quench} in the non-equilibrium quantum many-body community.
The relationship between conformal and laboratory time in the quench model is
\begin{equation}
    \eta(t) = \bar a \, t - \delta a \, t_0 \ln\left(\frac{a_\mathrm{f}\,\mathrm{e}^{t/t_0} + a_\mathrm{i}\,\mathrm{e}^{-t/t_0}}{a_\mathrm{f} + a_\mathrm{i}}\right)
    \label{eq:ConformalTimequench}
\end{equation}
Therein, we set $\eta(t = 0) = 0$ and introduce $\bar a = (a_\mathrm{f} + a_\mathrm{i}) / (2 a_\mathrm{f} a_\mathrm{i})$ and $\delta a = (a_\mathrm{f} - a_\mathrm{i}) / 2a_\mathrm{f} a_\mathrm{i}$.
The duration of the quench in conformal time is $\eta_0 = t_0 / a_\mathrm{f}$. Invoking this definition, we can plot both models as shown in \cref{fig:Models}.

As Parker noted in his other seminal paper \cite{Parker1976}, an exponential suppression of the particle spectrum at large wavenumbers resembles a thermal spectrum.
This resemblance becomes more exact the larger the ratio $a_\mathrm{f} / a_\mathrm{i}$ and is also the case for the Bernard-Duncan model \cite{BirrellDavies1982,Ford2021}.
From the scattering perspective, a thermal form of $N_k$ at temperature $T$ would require $\lvert r_+(k) \rvert^2 = \mathrm{exp}(-k/T)$ (cf. \cref{eq:ParticleNumber}).
Such an exactly thermal spectrum however would be infrared divergent, which is forbidden by the zero-energy resonance:
as we argued and explicitly checked in \cite{Schmidt2024}, the zero-energy resonance leads to a finite infrared limit of the particle number, $\lim_{k \to 0} N_k = (a_\mathrm{f} - a_\mathrm{i})^2 / (4 a_\mathrm{f} a_\mathrm{i})$, which indeed would look divergent if $a_\mathrm{f} \gg a_\mathrm{i}$.

Just an exponential decay of the spectrum is nevertheless allowed and satisfied by both the Bernard-Duncan model and the quench regardless of the ratio $a_\mathrm{f} / a_\mathrm{i}$.
We analyze this decay for both models in \cref{fig:ReflectionDecay} in the case of one e-fold, $a_\mathrm{f} / a_\mathrm{i} = \mathrm{e}$. 
The exponential decay can be understood from the scattering analogy via the identification $N_k = \lvert r_+(k) / t(k) \rvert^2$. 
The scattering amplitude of interest has the general form \cite{Chadan1989}
\begin{equation}
    \frac{r_\pm(k)}{t(k)}= (2\mathrm{i}k)^{-1} \int_{-\infty}^{\infty} \mathrm{d}\eta \, \mathrm{e}^{\mp \mathrm{i} k \eta} f_{\mp}(k,\eta) V(\eta).
\end{equation}
Due to the plane-wave asymptotics of the Jost solutions  $f_\pm(k,\eta)$, this relation becomes a Fourier-transform at high wavenumbers, 
\begin{equation}
    \begin{aligned}
            \frac{r_\pm(k)}{t(k)} \simeq (2\mathrm{i}k)^{-1} \hat V_\pm(2k), \quad
            \hat V_\pm(k) = \int_{-\infty}^{\infty} \mathrm{d} \eta \, V(\eta) e^{\mp \mathrm{i} k \eta}.
    \end{aligned}
\end{equation}
The exponential decay now follows from the Paley-Wiener theorem: 
if $V(\eta)$ is analytic for $\lvert \mathrm{Im} \, \eta \rvert < d$ we can shift the integration contour from the real axis to $\mathrm{Im} \, \eta = \pm d$.
This translation leads to an exponential damping factor $\lvert \hat V(k) \rvert \propto k^{m-1} \, \exp(- d \lvert k \rvert)$,
where $d$ is the distance from the real axis to the nearest singularity of $V(\eta)$ in the complex $\eta$-plane and $m$ is the order of the pole.
For instance, in the Bernard-Duncan model $a^2(\eta)$ has poles located at $\eta / \eta_0 = \mathrm{i} \pi (n + 1/2)$ from which the nearest is separated from the real axis by $d = \pi \eta_0 / 2$.
It does, however, also have a complex zero with the same imaginary part as the pole which is located at $\eta / \eta_0 = \ln(a_\mathrm{i}/a_\mathrm{f}) + \mathrm{i} \pi /2$.
Both are double poles of the Bernard-Duncan potential, hence
\begin{equation}
    \sqrt{N_k} \propto \mathrm{e}^{- \pi k \eta_0}.
    \label{eq:Nk_Decay}
\end{equation}
Repeating the same analysis for the quench model justifies the identification of $\eta_0 = t_0 / a_\mathrm{f}$:
the nearest singularity of the quench potential is the $\eta$-image of the pole of \eqref{eq:quench}, given by $\mathrm{i}\pi t_0 /2$.
The imaginary part of this image is $\mathrm{Im} \,  \eta(\mathrm{i}\pi t_0 /2) = \pi t_0 / (2 a_\mathrm{f})$ which can be identified with $\pi \eta_0 /2$.
It is again a second-order singularity of the potential, so \cref{eq:Nk_Decay} holds for the quench as well.

On a final note: According to the Paley-Wiener theorem, the exponential decay holds for any analytic potential $V(\eta)$ whereas a compactly supported potential can only lead to sub-exponential decay.
In a quantum simulation setting we deal with compactly supported potentials $V(\eta)$ due to on- and off-switching processes.

\subsection{Numerical Results}
\label{subsec:NumericalResults}

\begin{figure*}
    \includegraphics[width=\textwidth]{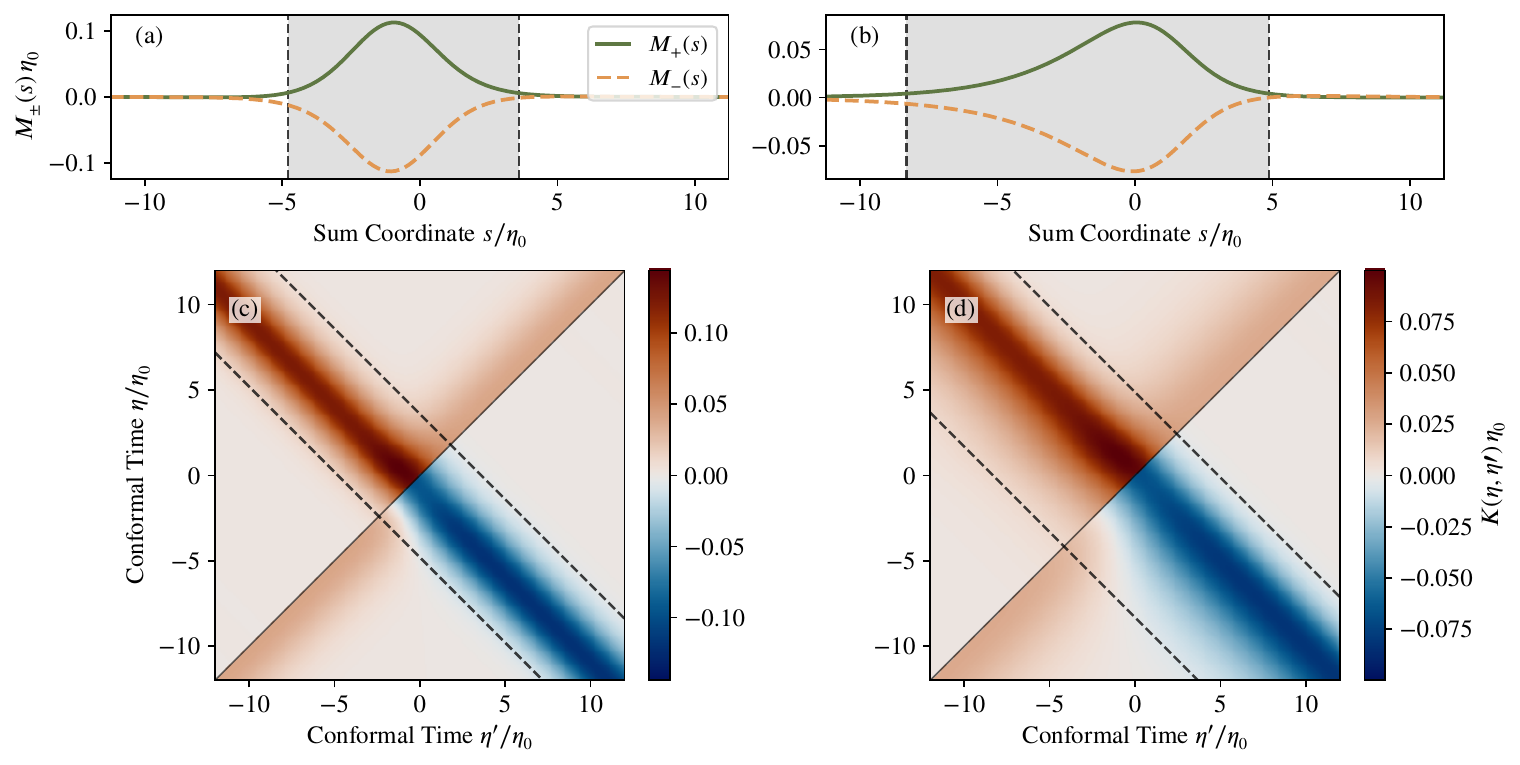}
    \caption{Marchenko functions $M_\pm(s)$ as a function of the sum coordinate $s = \eta + \eta'$ (top) and the resulting transformation kernels (bottom), for the Bernard-Duncan case (a,c) and the quench model (b,d).
    The upper triangle $\eta' > \eta$ shows $K_+$ and the lower triangle $\eta' < \eta$ shows $K_-$.
    Since \cref{eq:FaddeevMarchenkoUpper,eq:FaddeevMarchenkoLower} couple $K_\pm$ to $M_\pm$ through the sum of the arguments,
    $K_\pm(\eta,\eta') \simeq -M_\pm(\eta+\eta')$ at leading order, the kernel is constant along lines of constant $\eta + \eta'$ and its structure runs anti-diagonally at a perpendicular width of $\Delta s/\sqrt{2}$, where 
    $\Delta s$ is the length of the interval on which $|M_+|$ exceeds $5\%$ of its peak.}
    \label{fig:Marchenko}
\end{figure*}

\begin{figure}
    \includegraphics[width=\columnwidth]{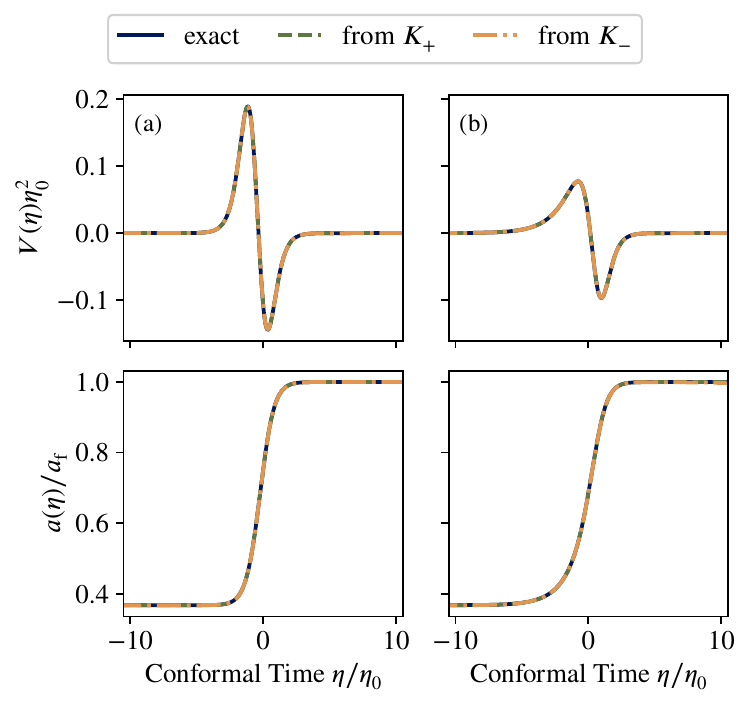}
    \caption{Reconstructed potential $V(\eta)\eta_0^2$ (top) and scale factor $a(\eta)/a_\mathrm{f}$ (bottom) from $K_\pm$ against the closed form for (a) Bernard-Duncan and (b) quench. Both branches agree to better than $0.02\%$ relative RMS [\cref{eq:RMS_relative}].}
    \label{fig:Reconstruction}
\end{figure}

After this short digression on key properties of the target models, let us invert them numerically. 
The Marchenko functions obtained from the FFT of the Numerov output are shown in the top row of \cref{fig:Marchenko}. 
The transformation kernels obtained from these $M_\pm$ by solving \cref{eq:FaddeevMarchenkoUpper,eq:FaddeevMarchenkoLower} 
via the Nyström-method are shown in the bottom row of the same figure, where the support of $M_\pm$ is carried over as a pair of dashed lines 
to highlight the width of the anti-diagonal band that forms as a result of the centered peak of $M_\pm$ (consider the caption of \cref{fig:Marchenko} for details).

We proceed by integrating the transformation kernel to reconstruct the scale-factor via \cref{eq:ScaleFactorReconstruction,eq:ZeroEnergyJost}
and show results obtained via both independent positive and negative half-strips in \cref{fig:Reconstruction}. Therein, the scattering potentials are shown as well for completeness.
\subsection{Robustness}

\begin{figure}[t]
    \includegraphics[width=\columnwidth]{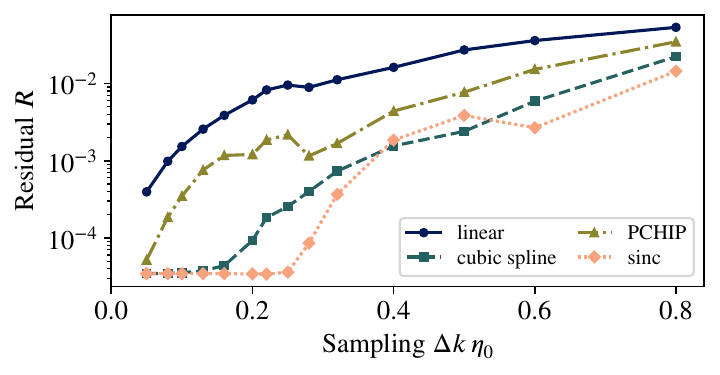}
    \caption{Relative RMS against the spacing $\Delta k$ of the measured wavenumber grid (Bernard-Duncan, $|\eta| \le 3\eta_0$, band
    $k_\mathrm{max}\eta_0 = 15$ held fixed), for four interpolations of
    $r_+(k)$ onto a finer, FFT-compatible grid.}
    \label{fig:kResolution}
\end{figure}

\begin{figure}[t]
    \includegraphics[width=\columnwidth]{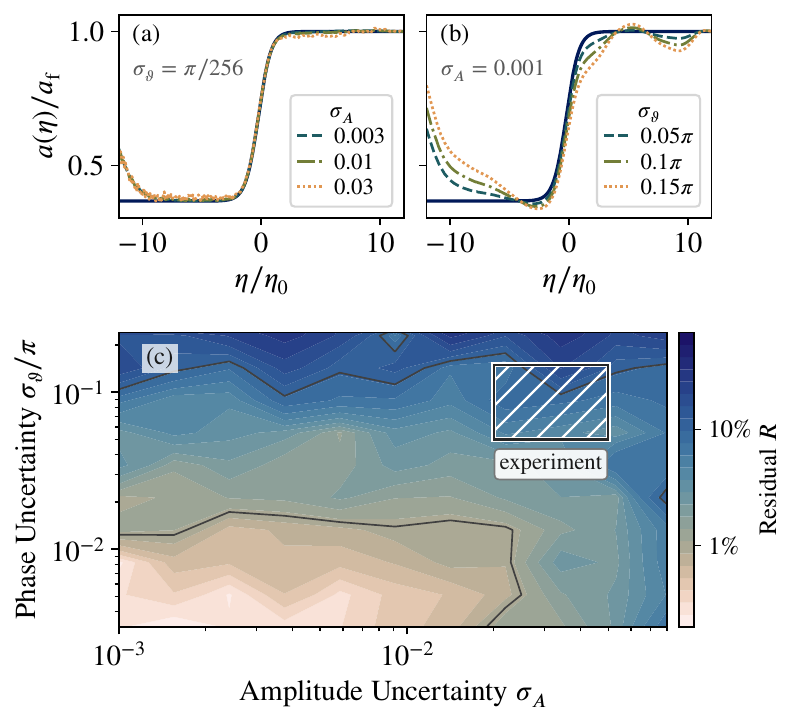}
    \caption{Reconstruction of the Bernard-Duncan scale factor $a(\eta)/a_\mathrm{f}$ via $K_+$ under measurement noise on (a) the oscillation amplitude and (b) the phase, each at negligible noise in the other.
    (c) RMS relative residual of $a(\eta)$ on $|\eta|\le 3\eta_0$ [\cref{eq:RMS_relative}] as a function of both uncertainties; the hatched box marks the experimental values of \cite{Sparn2024}.}
    \label{fig:Noise}
\end{figure}

Whether an inversion of this kind is stable against perturbations of the data is a
question with a literature of its own \cite{Marchenko1968, Bamberger1979, Aktosun1987stab, GraySymes1985}. 
In this section we ask two practical questions: how finely must the data be sampled and how accurately must each sample be known?
The following analysis focuses on the Bernard-Duncan model.

\subsubsection{Momentum space resolution}

The quantum field simulator for cosmology measures a finite set of discrete values for the scattering amplitudes as collected in \cref{eq:ParticleNumber,eq:Amplitude} at a certain momentum space resolution $\Delta k$.
It is therefore important to test the sensitivity of the reconstruction method to this resolution.
To that end $M_+(s)$ is built via a Fourier-sum of $r_+$ on a sparse, evenly-spaced $k$-grid of spacing $\Delta k$ which per construction returns an $M_+(s)$ that possesses artificial mirror copies beyond a window of width $2\pi / \Delta k$.
An interpolation of $r_+(k)$ on a finer, FFT-compatible grid can suppress these mirror copies, where we compare linear, shape-preserving (PCHIP), cubic-spline and sinc (Whittaker-Shannon) interpolation schemes.
We then vary the dimensionless resolution $\Delta k \eta_0$ while fixing the maximal wavenumber $k_\mathrm{max}\eta_0 = 15$ and the grid parameters.
The dependence of the solver's quality on $\Delta k \eta_0$ is quantified via
\begin{equation}
    R = \left[ \frac{1}{6\eta_0} \int_{-3 \eta_0}^{3 \eta_0} \mathrm{d}\eta \left( \frac{a_\mathrm{rec}^{(+)}(\eta)- a(\eta)}{a(\eta)} \right)^2 \right]^{1/2},
    \label{eq:RMS_relative}
\end{equation}
where $a^{(+)}_\mathrm{rec}(\eta)$ is the reconstructed scale-factor from the positive half-strip via $K_+$.
This error is shown in \cref{fig:kResolution} for the different interpolation schemes. 
We find that the sinc interpolation is best, as expected:
it would be exact for a band-limited $r_+(k)$ (compactly supported $M_+(s)$), but due to the exponential decay of $M_+(s)$ the sinc interpolation degrades once the tails of the Marchenko function $M_+(s)$ reach $\lvert s \rvert = \pi / \Delta k$. 

\subsubsection{Noise}

The preceding discussion isolates momentum space resolution effects, but a real measurement also carries uncertainties.
In the quantum field simulator for cosmology the scattering data are read off the density-fluctuation power spectrum,
that, at readout times, exhibits coherent oscillations which are fit via \cref{eq:Amplitude} (cf.\@ \cite{Viermann2022,Sparn2024}).
This leads to separately obtained amplitudes and phase-shifts that can be used to construct $r_+(k)$, given a proper model for the initial state.
We therefore perturb both the amplitude and phase with independent Gaussian errors $\sigma_A$ and $\sigma_\vartheta$ of fixed absolute size on every mode of the FFT-grid conjugate to the ROI. 

The influence of amplitude- (a) and phase-perturbations (b) on the target scale factor is shown in \cref{fig:Noise}.
The relative residual on the transition interval \eqref{eq:RMS_relative} (median over five noise realizations) is shown in \cref{fig:Noise} (c), where the hatched box corresponds to the error bars of Ref.~\cite{Sparn2024}, yielding $\sigma_A \approx 0.02$--$0.05$ and $\sigma_\vartheta \approx 0.05\,\pi$--$0.15\,\pi$.
A characteristic feature of the quantum simulation data is that phase-errors are high where the amplitude of the signal is low.
However, modeling that correlation explicitly through $\sigma_\vartheta(k) = [\sigma_\vartheta^2 + (\sigma_A / A_k)^2]^{1/2}$ leaves the residual essentially unchanged.
At that precision the scale factor is recovered to between about $5\%$ and $20\%$ with the phase as the binding constraint.

\section{Conclusion}
\label{sec:Conclusion}

We have inverted the description of (analog) cosmological particle production as a scattering problem \cite{Schmidt2024} based on the inverse problem in acoustics (see for instance \cite{DeiftTrubowitz1979}).
After having formulated the inverse problem for (analog) cosmology, we discussed to what extent one can recover the (analog) cosmological history (whose time-dependence generates the scattering landscape)
from the continuum scattering data which is experimentally accessible via the excitation power spectrum \cite{Viermann2022, Sparn2024}.
More concretely, the amplitude and phase of this spectrum can be used to assemble a complex reflection amplitude $r_+(k)$ 
from which one can infer the (analog) scalefactor $a(\eta)$ via solving the Faddeev-Marchenko equations.

We first established important structural properties of the scattering landscapes:
every non-singular, (analog) spacetime that evolves between two static epochs generates a potential with a zero-energy resonance $a^w(\eta)$, which is a bounded, nodeless solution to the mode equation at zero energy.
In this picture, cosmic (big-crunch--big-bang) singularities (zeros of $a(\eta)$) would correspond to nodes of this solution and lead to a bound-state of the scattering landscape.
The existence of a zero-energy resonance constitutes an exceptional case of scattering theory in the sense that the transmission amplitude does not vanish in the infrared, 
and needs oftentimes to be excluded by assumption or dealt with separately \cite{Chadan1989}.
We used this zero-energy resonance to derive an isospectrality between cosmological power spectra via a Darboux-transformation (which is also well-known from supersymmetric quantum mechanics \cite{Cooper2001} with the conformal Hubble rate corresponding to the superpotential):
For every non-singular expanding history $a(\eta)$ there is a contracting partner $\tilde a(\eta) = 1/a(\eta)$ whose potential is isospectral, producing a power spectrum that only differs by a $\pi$-shift in the phase.
The expansion- and contraction scenarios are therefore distinguished by the phase alone, which indicates that the phase is a robust discriminator among models (see \cite{Viermann2022} for a discussion with experimental data at hand).
Note that this duality relies on a well-defined conformal Hubble rate, such that singularities in $a$ and $\tilde a$ need to be avoided. 
Furthermore, this kind of duality map is known as a special limit transformation in the inverse-scattering literature \cite{Degasperis1987}.
We would like to stress that, although this relates to the expansion-contraction duality put forth by Wands \cite{Wands1999}, there is a significant difference:
Wands constructs the same scattering potential via different scalefactors, whereas the present duality includes different but \emph{isospectral} potentials.
Similar to the matter-bounce alternative based on Wands duality, the present isospectralities could be potentially relevant for primordial cosmology to construct further alternatives to inflation.

We progressed with stating the inverse cosmological particle production problem by inheritance of the inverse quantum scattering problem in one dimension with the restriction of no bound-state data if the scalefactor is regular.
For such regular histories the reconstruction is unique: the complex reflection amplitude entirely determines the potential \cite{Faddeev1964,DeiftTrubowitz1979}, and asymptotically static spacetimes admit no ambiguities in the reconstructed potential \cite{Degasperis1987} or its zero-energy Jost solution.
For the cosmological power spectrum, this means that the particle number $N_k$ and the coherence $\Delta N_k$ of the two-mode squeezed state need to be known for a unique reconstruction; or put differently, without the phase the reconstruction is non-unique, as the expansion-contraction duality shows.
The reconstruction can be carried out on two independent branches (the positive and negative half-strip in conformal time) which represents the asymptotic future and past, respectively.
Exact solutions to the inverse problem were obtained for two specific unimodal scattering potentials (the $\delta$-potential and the reflectionless Pöschl-Teller landscape) 
and used as benchmarks for the numerical scheme we have developed.
This included a numerical solution of the Fredholm integral equations of second kind via the Nyström method, where the quadrature rule was either Gauss-Legendre or piecewise linear, depending on the smoothness of the Marchenko function.
Each resulting transformation kernel (which encodes the scattering potential and all Jost solutions) agrees well with its closed form.
We were then able to apply the numerical algorithm to a Bernard-Duncan model and a quench of the analog scale-factor (logistic expansions in either $a^2(\eta)$ or $a(t)$) which followed via an integral of the transformation kernel.
We obtained the scale factor to better than $0.02\%$ in relative root-mean-square for both models and both branches.

Finally, we analyzed the robustness of the developed procedure in case of the Bernard-Duncan model regarding momentum-space resolution limits and measurement errors, which are both characteristic of an experimental situation.
A finite momentum space resolution leads to artificial mirror copies (known as aliasing in signal processing) which can be suppressed via interpolation, where a sinc (Whittaker-Shannon) interpolation scheme is optimal for quickly decaying Marchenko functions.
At the precision of Ref.~\cite{Sparn2024} a logistic scale factor can be recovered to between about five and twenty percent with the phase being the binding constraint.

The procedure has so far been applied to synthetic data; an interesting next step would be to use data from a quantum simulation.
Since the phase is the limiting input, it would be valuable to know how much can be recovered from the amplitude alone (the phase-retrieval problem \cite{Shechtman2015,Novikov2015}).
Other future prospects of this research include a generalization to the dispersive regime in which the modes experience a rainbow geometry that can be modeled via a wavenumber-dependent scattering potential \cite{Schmidt2026}.
Inverse scattering in the dispersive case has already been studied in the literature (see \cite[section IX.8]{Chadan1989} and references therein).
It would also be interesting to apply the concepts of this work to the two-dimensional scattering formulation of pair-production via analog black-holes or black hole superradiance and their quasi-normal modes.

Finally, a valuable contribution to discussion of early Universe physics could be to take a scale-invariant spectrum, build a scale-factor from it and then deduce which kind of early Universe histories are viable alternatives to inflation or bouncing cosmologies.

\begin{acknowledgments}
The authors thank Reinhard Meinel for illuminating discussions and Nico Dörk for valuable contributions to this work.
C. F. S. acknowledges support by the Studienstiftung des Deutschen Volkes and the Deutsche Forschungsgemeinschaft (DFG) under Grant No.~406116891 within the Research Training Group RTG 2522/1.
The graphics in this work were made using perceptually uniform color maps \cite{Crameri2023}.
\end{acknowledgments}

\clearpage
\appendix
\crefalias{section}{appendix}

\section{Isospectrality}
\label{sec:Isospectrality}
We closely follow \cite{Cooper2001}. Consider the Hamiltonian
\begin{equation}
    H_1 = - \dv[2]{\eta} + V_1(\eta)
    \label{eq:Hamiltonian}
\end{equation}
which describes one-dimensional scattering at a scattering potential $V_1(\eta)$.
Let us assume that it does not possess bound-states but may have a zero-energy resonance that is nodeless due to the absence of bound states.
Then one can factorize the Hamiltonian into $H_1 = A^\dagger A$ with 
\begin{equation}
    A = \dv{\eta} + W(\eta), \quad A^\dagger = -\dv{\eta} + W(\eta) 
    \label{eq:Factorization}
\end{equation}
where the quantity $W(\eta)$ is determined by the Riccati-equation
\begin{equation}
    V_1(\eta) = W^2(\eta) - W'(\eta). 
    \label{eq:Riccati}
\end{equation}
Let us now exchange the order of operators in the factorization $H_2 = A A^\dagger$, yielding the potential 
\begin{equation}
        V_2(\eta) = W^2(\eta) + W'(\eta)
\label{eq:V2}
\end{equation} 
whose scattering properties we want to compare to $V_1(\eta)$.

To that end consider an incident wave coming in from $\eta \to \infty$, propagating into negative $\eta$-direction, and scattering off the potentials $V_1$ and $V_2$ respectively. 
Here we consider a general mode dispersion $\sqrt{E_k}$.
In the asymptotic regions the respective wave-functions then have the form 
\begin{equation}
    \begin{aligned}
    \lim_{\eta \to - \infty} \psi_{1,2}(k,\eta) & = t_{1,2}(k) \mathrm{e}^{- \mathrm{i} \sqrt{E_k - W_-^2} \eta}, \\
    \lim_{\eta \to + \infty} \psi_{1,2}(k,\eta) &= \mathrm{e}^{- \mathrm{i} \sqrt{E_k - W_+^2} \eta} + r_{1,2}(k) \mathrm{e}^{\mathrm{i} \sqrt{E_k - W_+^2} \eta},
\end{aligned}
\label{eq:PlaneWaveAsymptotic}
\end{equation}
with $W_\pm = \lim_{\eta \to \pm \infty} W(\eta)$. 

In the described scattering problems, the factorizations of the Hamiltonians $H_1,H_2$ directly entail useful relations among their respective continuum eigenstates $\psi_k^{(1),(2)}$ with same eigenenergy $E_k$.
More concretely one finds that $A \psi_1(k,\eta)$ is an eigenstate of $H_2$, and vice versa that $A^\dagger \psi_2(k,\eta)$ is an eigenstate of $H_1$ with equal eigenenergy $E_k$ which directly entails that $\psi_2(k,\eta) = N A \psi_1(k,\eta)$, where $N$ is some normalization constant.  
Evaluating this relation for the asymptotic states defined in \cref{eq:PlaneWaveAsymptotic} yields a direct relation between the respective scattering amplitudes: 
\begin{equation}
    \begin{aligned}
        t_2(k) &= \frac{W_- - \mathrm{i} (E_k - W_-^2)^{1/2}}{W_+ - \mathrm{i} (E_k - W_+^2)^{1/2}} \, t_1(k) , \\
        r_2(k) &= \frac{W_+ +  \mathrm{i} (E_k - W_+^2)^{1/2}}{W_+ - \mathrm{i} (E_k - W_+^2)^{1/2}} \, r_1(k).
    \end{aligned}
\end{equation}
Since we consider an incoming left-mover, these expressions coincide with \cite{Cooper2001} upon exchanging $W_+$ with $W_-$.
Since the pre-factors are only phases one finds, $\big \lvert r_1(k) \big \rvert^2 =\big \lvert r_2(k) \big \rvert^2 $ and $ \big \lvert t_1(k) \big \rvert^2 = \big \lvert t_2(k) \big \rvert^2 $, i.e.\@ the respective probabilities of reflection and transmission are identical.  
In particular, in the case of a zero-energy resonance one has that $W_+ = W_- = 0$ and thus finds a simple relation between the amplitudes themselves, given by \cref{eq:IsospectralAmplitudes}.

\section{Scattering and Bound-State Data}
\label{app:Scat_Bound}
We recapitulate some basic facts based on \cite{Chadan1989}.
\subsection{Scattering data}
Since the span of $\{f_\pm(k,\eta), f_\pm(-k,\eta)\}$ is a fundamental system for the Schrödinger equation, we can write
\begin{equation} 
    f_\mp(k,\eta) = \frac{1}{t(k)} f_\pm(-k,\eta) +  \frac{r_\pm(k)}{t(k)} f_\pm(k,\eta), \label{eq:JostRelation}
\end{equation}
where the coefficients can be related to the transmission amplitude $t(k)$ and the reflection amplitude $r_\pm(k)$ by comparison to the scattering solutions.
These can be extracted directly via the Wronskians,
\begin{align}
    \frac{1}{t(k)} &= \frac{1}{2\mathrm{i}k} \mathrm{Wr}[f_-(k,\eta), f_+(k,\eta)], \\ 
    \frac{r_+(k)}{t(k)} &= - \frac{1}{2 \mathrm{i}k} \mathrm{Wr}[f_-(k,\eta), f_+(-k,\eta)], \\
    \frac{r_-(k)}{t(k)} &= \frac{1}{2 \mathrm{i}k} \mathrm{Wr}[f_+(k,\eta), f_-(-k,\eta)].
\end{align}
Due to the unitarity of $S$, one has the relations
\begin{align}
    r_+(k) t(-k) + r_-(-k) t(k) &= 0, \label{eq:ForwardBackwardReflectionRel}\\
    r_+(k) r_+(-k) + t(k) t(-k) &= r_-(k) r_-(-k) + t(k) t(-k) = 1,
\end{align}
and for $k \in \mathbb{R}$, 
\begin{equation}
    r_\pm(-k) = r_\pm(k)^*, \quad t(-k) = t(k)^*.
\end{equation}

\subsection{Bound state data}
The finitely many bound-states at energies $- \kappa_n^2$ correspond to simple poles of $t(k)$ at $\mathrm{i}\kappa_n$ where
$\kappa_n > 0$ for $n \in \lbrace 0,\ldots, N-1 \rbrace$.
By virtue of \cref{eq:JostRelation}, when evaluated at $\mathrm{i}\kappa_n$, the Jost solutions $f_\pm(\mathrm{i}\kappa_n,\eta)$ are directly proportional to the $L^2$-normalized bound-state wavefunction $\psi_n(\eta)$, i.e.\@
\begin{equation}
    f_-(\mathrm{i}\kappa_n, \eta) = C_n^- \psi_n(\eta), \quad f_+(\mathrm{i} \kappa_n, \eta) = C_n^+ \psi_n(\eta),
\end{equation}
where the real proportionality coefficients follow from 
\begin{equation}
    (C_n^{\pm})^{2} = \int_{-\infty}^{\infty} \mathrm{d} \eta \, f_\pm(\mathrm{i}\kappa_n,\eta)^2,
    \label{eq:BoundStateNormalizationJost}
\end{equation}
and can be directly expressed in terms of the scattering amplitudes via
\begin{equation}
    (C_n^{\pm})^{-2} = - \mathrm{i} b_\pm(\mathrm{i} \kappa_n) \, \left(\frac{\partial a}{\partial k}\right)\bigg \lvert_{k = \mathrm{i} \kappa_n}^{-1},
\end{equation}
where 
\begin{equation}
    b_\pm(k) = \frac{r_\pm(k)}{t(k)}, \quad a(k) = t(k)^{-1}.
    \label{eq:abDefinition}
\end{equation}
Note that these ratios of scattering coefficients are connected to the bound-state parameters via
\begin{equation}
    b_\pm(\mathrm{i}\kappa_n) = C_n^\mp / C_n^\pm, \quad a(\mathrm{i}\kappa_n) = 0.
\end{equation}
Another useful identity concerns the residue of the transmission coefficient at its pole, $\mathrm{Res}(t, \mathrm{i}\kappa_n) =  \mathrm{i} / (C_n^+ C_n^-)$.

\begin{widetext}
\section{Derivation of Faddeev-Marchenko equation}
\label{sec:FM_Derivation}
See also \cite{Sabatier1983}.
First and foremost let us introduce the auxiliary functions 
\begin{equation}
    F_\pm(k,\eta) = \exp(\mp \mathrm{i} k \eta) f_\pm(k,\eta)
    \label{eq:AuxiliaryFunctions}
\end{equation}
in terms of which the Levin representation \eqref{eq:LevinRepJost} of the Jost functions reads
\begin{equation}
    F_\pm(k,\eta) = 1 + \int_{-\infty}^{\infty} \mathrm{d}\eta' \, \Theta[\pm (\eta' - \eta)] \, \exp[\pm \mathrm{i} k (\eta' - \eta)] K_\pm(\eta, \eta') .
    \label{eq:LevinF}
\end{equation}
Since $F_\pm(k,\eta)-1$ are holomorphic functions of $k$ for $\mathrm{Im} \, k \geq 0$ which decay as $\mathcal{O}(\lvert k \rvert^{-1})$ for $\lvert k \rvert \to \infty$, 
we can invoke the Cauchy integral formula
\begin{equation}
    F_\pm(k,\eta) - 1 = \frac{1}{2\pi \mathrm{i}} \lim_{\epsilon \to 0^+}  \int_{-\infty}^{\infty}  \mathrm{d} k' \frac{F_\pm(k',\eta) - 1}{k' - k - \mathrm{i}\epsilon}.
\end{equation}
Performing the integral substitution $k' \to - k'$ and using \cref{eq:JostRelation} one can write the right-hand-side as
\begin{equation}
    F_\pm(k,\eta) - 1  =  \lim_{\epsilon \to 0^+}  \int_{-\infty}^{\infty} \frac{ \mathrm{d} k'}{2\pi \mathrm{i}}  \left[ \frac{1 - t(k') F_\mp(k',\eta)}{k' + k + \mathrm{i}\epsilon} + \frac{r_\pm(k') F_\pm(k',\eta) \exp(\pm 2 \mathrm{i} k' \eta)}{k' + k + \mathrm{i}\epsilon} \right].
    \label{eq:FM_Deriv_StartingPoint}
\end{equation}
The integral involving the transmission amplitude can be further evaluated in terms of the residue theorem.
Closing in the upper half-plane we find
\begin{equation}
 \lim_{\epsilon \to 0^+}  \int_{-\infty}^{\infty} \frac{ \mathrm{d} k'}{2\pi \mathrm{i}} \frac{1 - t(k') F_\mp(k',\eta)}{k' + k + \mathrm{i}\epsilon} = - \mathrm{i} \sum_{n=0}^{N-1} \frac{\exp(\mp 2 \kappa_n \eta)}{k + \mathrm{i} \kappa_n} (C_n^\pm)^{-2} F_\pm(\mathrm{i}\kappa_n,\eta),
 \label{eq:FM_Deriv_TransmissionContribution}
\end{equation}
where we used the definition of the auxiliary functions \eqref{eq:AuxiliaryFunctions} and the identities stated in the discussion around \cref{eq:BoundStateNormalizationJost}.

As a next step, we perform two distinct Fourier transforms of \cref{eq:FM_Deriv_StartingPoint} after inserting \cref{eq:FM_Deriv_TransmissionContribution}.
More concretely, again employing the residue theorem yields
\begin{align}
    \int_{-\infty}^\infty \frac{\mathrm{d}k}{2\pi} \, \mathrm{e}^{- \mathrm{i} k (\eta' - \eta)} \, [F_+(k, \eta) - 1]
    &= - \mathrm{i} \sum_{n=0}^{N-1} \mathrm{e}^{- \kappa_n (\eta + \eta')} \left(C_n^+ \right)^{-2} F_+(\mathrm{i} \kappa_n, \eta)
    - \mathrm{i} \int_{-\infty}^{\infty} \frac{ \mathrm{d} k'}{2\pi} \mathrm{e}^{\mathrm{i} k'(\eta + \eta')}  \,  r_+(k') F_+(-k',\eta) \quad (\eta' > \eta),
    \label{eq:FM_Deriv_FourierResultUpper} \\
    \int_{-\infty}^\infty \frac{\mathrm{d}k}{2\pi} \, \mathrm{e}^{ \mathrm{i} k (\eta' - \eta)} \, [F_-(k, \eta) - 1]
    &= - \mathrm{i} \sum_{n=0}^{N-1} \mathrm{e}^{\kappa_n (\eta + \eta')} \left(C_n^- \right)^{-2} F_-(\mathrm{i} \kappa_n, \eta)
    - \mathrm{i} \int_{-\infty}^{\infty} \frac{ \mathrm{d} k'}{2\pi} \mathrm{e}^{- \mathrm{i} k'(\eta + \eta')}  \,  r_-(k') F_-(-k',\eta) \quad (\eta' < \eta),
      \label{eq:FM_Deriv_FourierResultLower}
\end{align}
where we close in the lower half plane.
The final step is to insert the Levin representation \eqref{eq:LevinF} for $F_\pm$ on both sides of \cref{eq:FM_Deriv_FourierResultUpper,eq:FM_Deriv_FourierResultLower}, respectively.
One finds
\begin{equation}
    \begin{aligned}
            K_+(\eta, \eta') =&
            - \sum_{n=0}^{N-1} \exp[-\kappa_n (\eta + \eta')] \left(C_n^+\right)^{-2} - \int_{-\infty}^{\infty} \frac{ \mathrm{d} k}{2\pi} \exp[\mathrm{i} k(\eta + \eta')]  \,  r_+(k) \\
            &- \int_{\eta}^{\infty} \mathrm{d} \eta'' \, K_+(\eta, \eta'') \left[ \sum_{n=0}^{N-1} \left(C_n^+\right)^{-2} \exp(-\kappa_n (\eta'' + \eta')) + \int_{-\infty}^{\infty} \frac{ \mathrm{d} k}{2\pi} \exp(\mathrm{i} k(\eta'' + \eta'))  \,  r_+(k) \right],
    \end{aligned}
\end{equation}
for $\eta' > \eta$, and
\begin{equation}
    \begin{aligned}
            K_-(\eta, \eta') =&
            - \sum_{n=0}^{N-1}  \exp[\kappa_n (\eta + \eta')] \left(C_n^-\right)^{-2}- \int_{-\infty}^{\infty} \frac{\mathrm{d} k}{2\pi} \exp[-\mathrm{i} k(\eta + \eta')]  \,  r_-(k) \\
            &- \int_{-\infty}^{\eta} \mathrm{d} \eta''  K_-(\eta, \eta'') \left[ \sum_{n=0}^{N-1} \left(C_n^-\right)^{-2} \exp(\kappa_n (\eta'' + \eta')) + \int_{-\infty}^{\infty} \frac{ \mathrm{d} k}{2\pi} \exp(-\mathrm{i} k(\eta'' + \eta'))  \,  r_-(k) \right],
    \end{aligned}
\end{equation}
for $\eta' < \eta$. 
\end{widetext}

\section{Potentials with compact support}
\label{subsec:PotentialCompact_Sol}
If the potential vanishes outside a compact region, $V(\eta) = 0$ for $\eta \notin[0,\Delta \eta]$, the Jost solutions are \cite{Portinari1967}
\begin{align}
    f_+(k,\eta) &= t(k)^{-1} \left(\mathrm{e}^{\mathrm{i} k \eta} + r_-(k) \, \mathrm{e}^{-\mathrm{i} k \eta}\right) \quad &(\eta < 0), \label{eq:FreeJostLower} \\
    f_-(k,\eta) &= t(k)^{-1} \left(\mathrm{e}^{-\mathrm{i} k \eta} + r_+(k) \, \mathrm{e}^{\mathrm{i} k \eta}\right) \quad &(\eta > \Delta \eta). \label{eq:FreeJostUpper} 
\end{align}
It is then possible to show that the Marchenko functions satisfy $M_+(\eta) = 0$ for $\eta > 2 \Delta \eta$ and $M_-(\eta) = 0$ for $\eta < 0$.
Furthermore it is possible to show that $V(\eta) = 0$ for $\eta \notin [0, \Delta \eta]$ implies that \cite{Portinari1967}
\begin{equation}
    \int_{-\infty}^\infty \mathrm{d}k \, k \frac{r_-(k)}{t(k)} \, \mathrm{e}^{-2\mathrm{i} k \eta}= 0
\end{equation}
for $\eta \notin [0, \Delta \eta]$.
Moreover, invoking \cref{eq:FreeJostLower,eq:FreeJostUpper}, one can compute the transformation kernel directly \cite{Portinari1967}.
One finds
\begin{widetext}
    \begin{align}
    K_+(\eta, \eta')
    = - \int_{-\infty}^\infty \frac{\mathrm{d}k}{2\pi}  \, \left[ \frac{r_+(k)}{t(k)} \mathrm{e}^{\mathrm{i} k (\eta + \eta')}  + \frac{r_+(k) \, r_-(k)}{t(k)} \mathrm{e}^{-\mathrm{i} k (\eta - \eta')} \right]
    - \sum_n (C_n^+ C_n^-)^{-1} \mathrm{e}^{- \kappa_n (\eta' - \eta)},
     \label{eq:KernelPositive}
\end{align}
for $\eta' > \eta$, if $V(\eta) = 0$ for all $\eta < 0$.
Similarly one has
\begin{align}
    K_-(\eta, \eta')
    = - \int_{-\infty}^\infty \frac{\mathrm{d}k}{2\pi}  \, \left[ \frac{r_-(k)}{t(k)} \mathrm{e}^{-\mathrm{i} k (\eta + \eta')}  + \frac{r_+(k) \, r_-(k)}{t(k)} \mathrm{e}^{\mathrm{i} k (\eta - \eta')} \right]
    - \sum_n (C_n^+ C_n^-)^{-1} \mathrm{e}^{\kappa_n (\eta' - \eta)}, \label{eq:KernelNegative}
\end{align}
for $\eta' < \eta$, if $V(\eta) = 0$ for all $\eta > \Delta \eta$.
\end{widetext}
Note that, in order to arrive at \cref{eq:KernelPositive,eq:KernelNegative}, we straightforwardly extended the calculation performed in \cite{Portinari1967} to the possible presence of bound states.
Interestingly, these do not enter the diagonal kernel derivative of the transformation kernels, i.e.\@
\begin{align}
\dv{\eta} K_+(\eta, \eta) &= -\frac{\mathrm{i}}{\pi} \int_{-\infty}^\infty \mathrm{d}k \, k \, \frac{r_+(k)}{t(k)} \mathrm{e}^{2 \mathrm{i} k \eta} \quad (\eta < 0), \label{eq:KernelDerivativePositive}  \\
\dv{\eta} K_-(\eta, \eta) &= \frac{\mathrm{i}}{\pi} \int_{-\infty}^\infty \mathrm{d}k \, k \, \frac{r_-(k)}{t(k)} \mathrm{e}^{-2 \mathrm{i} k \eta} \quad (\eta > \Delta \eta).  \label{eq:KernelDerivativeNegative}  
\end{align}

\section{Sabatier's algorithm for rational scattering amplitudes}
\label{subsec:RationalScat_Sol}
In \cite{Sabatier1983} it was shown that inverse quantum scattering in one dimension can be solved if $r(k)$ and $t(k)$ are rational fractions that satisfy additional boundary constraints. 
To illustrate a resolution of the inverse problem, we summarize the algorithm here.
The transmission amplitude needs to satisfy: 
\begin{enumerate}
    \item[(i)] $\lim\limits_{\lvert k \rvert \to \infty} t(k) = 1$.
    \item[(ii)]  The only zero of $t(k)$ in the closed upper half plane is at $k = 0$.
    \item[(iii)] The only poles of $t(k)$ in the upper half plane lie on the purely imaginary axis.
\end{enumerate}  
The reflection amplitudes needs to satisfy:
\begin{enumerate}
    \item[(i)]  $r_\pm(k)$ decays at least as $\lvert k \rvert^{-1}$ as $\lvert k \rvert \to \infty$.
    \item[(ii)] $\lim\limits_{\lvert k \rvert \to 0} r_\pm(k) = -1 $.
\end{enumerate}
Note that these conditions exclude a zero-energy resonance.
If we assume that all poles of $r_\pm(k)$ in the upper half plane are simple and contain them in the set $\mathcal{P}^\pm$ whereas all poles of $t(k)$ are the aforementioned bound-states $\mathrm{i} \kappa_n$,
we can express the Marchenko functions as 
\begin{equation}
    M_+(\eta) = \mathrm{i} \sum_{p_+ \in \mathcal{P}^+} \Res(r_+, p_+) \exp(\mathrm{i} p_+ \eta) + \sum_{n=0}^{N-1} \frac{\exp(-\kappa_n \eta)}{(C_n^+)^{2}} ,
\end{equation}
for $\eta > 0$, and 
\begin{equation}
    M_-(\eta) = \mathrm{i} \sum_{p_- \in \mathcal{P}^-} \Res(r_-, p_-) \exp(-\mathrm{i} p_- \eta) + \sum_{n=0}^{N-1} \frac{\exp(\kappa_n \eta)}{(C_n^-)^{2}},
\end{equation}
for $\eta < 0$.
Here, we close in the upper half plane, hence requiring $\eta > 0$ (respectively $\eta < 0)$ in the upper (respectively lower) case.
\begin{widetext}
Due to the similarity between the two summands, it is suggestive to unite them via the sets
\begin{align}
    \mathcal{Z}^\pm &\equiv \mathcal{P}^\pm \cup \lbrace \mathrm{i} \kappa_0, \ldots, \mathrm{i} \kappa_{N-1} \rbrace \\
    \Rho^\pm &\equiv \lbrace \Res(r_\pm, p^\pm) | p^\pm \in \mathcal{P}^\pm \rbrace \cup  \big \lbrace - \mathrm{i} \left(C_0^\pm\right)^{-2}, \ldots , - \mathrm{i} \left(C_{N-1}^\pm\right)^{-2} \big \rbrace
\end{align}
in order to write
\begin{equation}
    M_\pm(\eta) = \mathrm{i} \sum_n \rho_n^\pm \exp(\pm \mathrm{i} z_n^\pm \eta), 
\end{equation}
where $\rho_n^\pm \in \Rho^\pm$ and $z_n^\pm \in \mathcal{Z}^\pm$.
The FMEs then read
\begin{align}
    K_+(\eta,\eta') + \mathrm{i} \sum_{n} \rho_n^+ \exp[\mathrm{i} z_n^+ (\eta + \eta')] + \mathrm{i} \sum_{n} \rho_n^+ \exp(\mathrm{i} z_n^+\eta') \int_\eta^{\infty} \dd \eta'' \exp(\mathrm{i} z_n^+\eta'') \, K_+(\eta,\eta'') = 0 \qquad (&\eta' > \eta > 0), \\
    K_-(\eta,\eta') + \mathrm{i} \sum_{n} \rho_n^- \exp[-\mathrm{i} z_n^- (\eta + \eta')] + \mathrm{i} \sum_{n} \rho_n^- \exp(-\mathrm{i} z_n^-\eta') \int_{-\infty}^{\eta} \dd \eta'' \exp(-\mathrm{i} z_n^-\eta'') \, K_-(\eta,\eta'') = 0 \qquad (&\eta' < \eta < 0),
\end{align}
\end{widetext}
and are solved by a separable (or degenerate) transformation kernel 
\begin{equation}
    K_\pm(\eta, \eta') = \sum_{n} k_n^\pm(\eta) \exp(\pm \mathrm{i} z_n^\pm \eta').
\end{equation}
Indeed, inserting this Ansatz, the integration becomes elementary and one finds the linear system of algebraic equations 
\begin{equation}
\sum_m A^\pm_{nm}(\eta) k_m^\pm(\eta) + \mathrm{i} \rho_n^\pm \exp(\pm \mathrm{i} z_n^\pm \eta) = 0
\end{equation}
with the matrix elements 
\begin{equation}
    A^{\pm}_{nm} = \delta_{nm} - \frac{\rho_n^\pm}{z_n^\pm + z_m^\pm} \exp \left[\pm \mathrm{i} (z_n^\pm + z_m^\pm) \eta\right]
\end{equation}
upon using the linear independence of the functions $\exp(\pm \mathrm{i} z_n^\pm \eta')$.
Inserting the solution
\begin{equation}
  k_m^\pm(\eta) = - \sum_n \mathrm{i} (A^\pm)^{-1}_{mn}(\eta) \rho_n^\pm \exp(\pm \mathrm{i} z_n^\pm \eta)  
\end{equation}
back into the transformation kernel and evaluating it on the diagonal we find
\begin{equation}
    \begin{aligned}
    K_+(\eta, \eta) &= - \mathrm{i} \sum_{m,n} (A^+)^{-1}_{mn} \rho_n^+ \exp[\mathrm{i} (z_n^+ + z_m^+)\eta]  \quad (\eta > 0) \\
    K_-(\eta, \eta) &= - \mathrm{i} \sum_{m,n} (A^-)^{-1}_{mn} \rho_n^- \exp[-\mathrm{i} (z_n^- + z_m^-)\eta] \quad (\eta < 0)
    \end{aligned}
\end{equation}
which can be brought into the convenient form
\begin{equation}
    \begin{aligned}
        K_+(\eta, \eta) &=  \frac{\mathrm{d}}{\mathrm{d}\eta} \log \det A^+ \quad (\eta > 0) \\
        K_-(\eta, \eta) &= - \frac{\mathrm{d}}{\mathrm{d}\eta} \log \det A^- \quad (\eta < 0)
    \end{aligned}
\end{equation}
where we used that 
\begin{equation}
    \dv{\eta} A^\pm_{nm}(\eta) = \mp \mathrm{i} \rho_n^\pm  \exp[\pm \mathrm{i} (z_n^\pm + z_m^\pm) \eta]
\end{equation}
According to the Deift-Trubowitz conditions \cite{DeiftTrubowitz1979} (which are guarenteed by the above conditions on the reflection- and transmission amplitude)
this solution is unique on both the negative half strip ($\eta < 0$) and the positive half strip ($\eta > 0$). Hence,
\begin{equation}
    \begin{aligned}
            V(\eta) &= - 2 \log''[\det A^\pm(\eta)] \\
            &= 2 \left(\frac{[\det A^\pm(\eta)]'}{\det A^\pm(\eta)}\right)^2 - 2 \frac{[\det A^\pm(\eta)]''}{\det A^\pm(\eta)}
    \end{aligned}
\end{equation}
is the unique solution to the inverse problem.

\clearpage 
\bibliography{main}
\bibliographystyle{apsrev4-2} 

\end{document}